\documentclass[nopreprint,a4paper,floatfix,noshowpacs,twocolumn]{revtex4}

\usepackage{graphicx}

\usepackage{extarrows}  
\usepackage{amsmath,amsfonts,amssymb,dsfont}

\usepackage{fancyhdr}
\usepackage{microtype}

\usepackage[sf,small,raggedright]{titlesec}

\def\theequation{\arabic{section}.\arabic{equation}}

	\newcommand{\MittagLeffler}[3]{\ensuremath{E_{#1,#2} \big(#3\big)}}
	\newcommand{\MittagLefflerAlphaBetaT}{\MittagLeffler{a}{b}{t}}

	\newcommand{\C}{\ensuremath{\mathbb{C}}}

	\newcommand{\Fouriertransform}[1]{\ensuremath{\mathcal F}\left\{{#1}\right\}}
	
	\newcommand{\Fourierbackfourth}{\ensuremath{\xleftrightarrow{\ \textstyle\mathcal F\ }}}

	\newcommand{\dd}{\ensuremath{\text{d}}}

	\newcommand{\taue}{\ensuremath{\tau_\epsilon}}
	\newcommand{\taus}{\ensuremath{\tau_\sigma}}
	\newcommand{\taun}{\ensuremath{\tau_\nu}}
	\newcommand{\kappaz}{\ensuremath{\kappa_0}}
	\newcommand{\kappan}{\ensuremath{\kappa_\nu}}
	\newcommand{\kappao}{\ensuremath{\kappa(\omega)}}

	\newcommand{\eg}{\mbox{{e.g.}}}
	
\begin{document}
\title%
{{\small SURVEY PAPER}\\\smallskip On a Fractional Zener Elastic Wave Equation}%
\author{Sven Peter N\"asholm}\email{svenpn@i{f}i.uio.no}\affiliation{Department of Informatics, University of Oslo, P.~O.~Box 1080, NO--0316 Oslo, Norway} 
\author{Sverre Holm}\affiliation{Department of Informatics, University of Oslo, P.~O.~Box 1080, NO--0316 Oslo, Norway}


\begin{abstract} 
\noindent This survey concerns a causal elastic wave equation which implies frequency power-law attenuation. %
The wave equation can be derived from a fractional Zener stress-strain relation plus linearized conservation of mass and momentum. %
A connection between this four-parameter fractional wave equation and a physically well established multiple relaxation acoustical wave equation is reviewed. %
The fractional Zener wave equation implies three distinct attenuation power-law regimes and a continuous distribution of compressibility contributions which also has power-law regimes. %
Furthermore it is underlined that these wave equation considerations are tightly connected to the representation of the fractional Zener stress-strain relation, which includes the \emph{spring-pot} viscoelastic element, and by a Maxwell--Wiechert model of conventional springs and dashpots. %
A purpose of the paper is to make available recently published results on fractional calculus modeling in the field of acoustics and elastography, with special focus on medical applications. 


\bigskip
\noindent\texttt{The peer-reviewed version of this paper is now published in Fract.\ Calc.\ Appl.\ Anal.\ Vol.\ 16, No 1 (2013), pp.\ 26-50, DOI: 10.2478/s13540-013--0003-1, which is a Special Issue for FDA'12. It will be available at http://link.springer.com/journal/13540}

\smallskip\noindent\texttt{The current document is an e-print which differs in e.g.\ pagination, reference numbering, and typographic detail.}
\end{abstract}



\maketitle
\section{Introduction\label{Sec:1}}

This paper investigates connections between fractional viscoelasticity, fractional wave equations, causal models, and power-law attenuation within the framework of elastic wave modeling. %
Paying extra attention to medical imaging applications, we intend to convey to the fractional calculus community information on developments related to time-fractional elastic wave equations. %
The paper expands upon the conference proceeding paper \cite{Nasholm2012China} and the J.\ Acoust.\ Soc.\ Am.\ papers \cite{Holm2011, Nasholm2011}. %

In Section \ref{sec:powerlaw}, the frequently encountered power-law attenuation nature of elastic waves in complex media is considered, especially in medical imaging. %

In Section \ref{sec:fractionalzener}, a fractional generalization of the Zener viscoelastic model is reviewed. Parameter value restrictions to keep the model thermodynamically admissible are discussed and %
experimental evidences are listed. %
The section also demonstrates the important achievement that the fractional Zener model is equivalent to a Maxwell--Wiechert representation consisting of a set of conventional springs and dashpots. Also physically underlying principles which lead to fractional viscoelasticity are reflected upon. %

Subsequently, Section \ref{sec:wave_equations} derives a fractional Zener wave equation from conservation laws and the fractional Zener constitutive stress-strain model. %
Section \ref{sec:properties} then analyzes properties of this wave equation. A connection to the widely acknowledged Nachman--Smith--Waag wave equation for acoustic propagation with relaxation losses \cite{Nachman1990} is demonstrated, attenuation and finite phase speed power-law regimes are evaluated, and causality conditions are considered. 

The final section provides conclusions and discussions including parallels to similar models in other fields, such as electromagnetics. %

%
%
\section{Power-law attenuation in complex media\label{sec:powerlaw}}
%
%

Elastic wave attenuation in complex media such as biological tissue, polymers, rocks, and rubber often follows a frequency power law: 
\begin{align}
	\alpha_k(\omega) \propto \omega^\eta,
\end{align} 
with the exponent $\eta \in [0,2]$. %
Such power-laws can be valid over many frequency decades and examples are found all the way from infrasound to ultrasound \cite{Duck1990}. %
See \eg{} Fig.~1 in \cite{Szabo00} which visualizes experimentally established power-law attenuation examples for both shear and compressional waves. %
As summarized in \cite{Holm2010}, compressional wave attenuation in biological tissue commonly manifests a power-law exponent $\eta \in[1,2]$, while for shear waves in tissue one often finds $\eta \in[0,1]$.

\subsection{Power-law attenuation in medical imaging}

The establishment of accurate wave-propagation models in power-law lossy media is important for applications where broadband waves are utilized. %
Below we list examples of such applications.

In photoacoustic imaging and tomography, laser pulses are delivered into tissue where an associated thermoelastic expansion induces a wideband ultrasound emission which is used for image reconstruction \cite{beard2011biomedical,kowar2012attenuation, roitner2012experimental, treeby2010reconstruction}. %

Magnetic resonance (MR) elastography is a another means to estimate soft tissue stiffness \eg{} in the liver or the breast. This method utilizes the propagation of shear waves which are monitored using MR imaging \cite{2012ehmanreview, muthupillai1995science, papazoglou2012multifrequency, Sinkus2000, Sinkus2007, sinkus2012review, yasar2012wideband}. %
Doppler techniques may also be applied to monitor the frequency-dependent tissue shear wave properties \cite{barry2012shearwavesispersion}. %

A related elastography method is acoustic radiation force imaging (ARFI) techniques, where tissue is deformed by a short compressional pulse which creates a propagating shear wave. The displacement due to the shear wave as measured by ultrasound is then used to quantify the tissue's mechanical properties \eg{} by estimating the shear wave propagation speed \cite{bercoff2004supersonic, Chen2004, palmeri2011acoustic}. %

\subsection{Modeling power-law attenuation}

For acoustic modeling, time-fractional derivative wave equations have been shown to imply power-law attenuation over wide frequency bands \cite{Holm2011, Holm2010}. %
As further described in Section \ref{sec:wave_equations}, fractional wave equations can be obtained from linearized conservations of mass and momentum in combination with time-fractional constitutive relations connecting stress and strain. %
Related linear wave-propagation simulations are reported \eg{} in \cite{Wismer1995, Liebler2004, Wismer06, Caputo2011}. %
For waves at high amplitudes, non-linear effects need to be considered. Such models were developed in \cite{Prieur2011, Prieur2012} where the fractional Kelvin--Voigt constitutive equation was applied. See also the related recent paper \cite{treeby2012modeling}. %

From a numerical modeling point of view when considering the low-frequency or small-attenuation regimes of a power-law attenuating medium, wave equations with a d'Alambertian part and time-fractional terms may conveniently be converted into space-fractional Laplacian models. This can be beneficial due to reduced time-signal storage needed in the propagation simulation steps \cite{Chen03, Treeby2010, Carcione2010, treeby2012modeling}. Special care needs to be taken to ensure causality of the resulting models. %
The authors are not aware of similar valid conversions between fractional temporal derivatives and fractional Laplacians that are applicable for attenuation with $\eta < 1$ or within a high-frequency regime. %

The multiple relaxation mechanism framework of \cite{Nachman1990} is widely considered as adequate for acoustic wave %
modeling in lossy complex media like those encountered in medical ultrasound. %
It relies on thermodynamics and first principles of acoustics. The corresponding wave equation for $N$ relaxation mechanisms is a causal partial differential equation with its highest time derivative order $N+2$. We denote this the Nachman--Smith--Waag (NSW) model. %
In order to make the discrete NSW model attenuation adequately follow $\omega^\eta$, either the valid frequency band must be narrow, or the number of assumed mechanisms $N$ must be large thus inferring a partial differential equation of high order. %

On the other hand, for a certain continuous distribution of relaxation mechanisms, the NSW and fractional Zener descriptions are equal and power-law attenuation is attained, as further reviewed in Section \ref{sec:link_to_nachman}. 

Band-limited fits to power-law acoustic attenuation for relaxation models with $N=2$ and $3$ are exemplified by \cite{Tabei2003, Yang2005}. In the latter, one of the mechanisms is assumed to be of very high relaxation frequency, thus representing a thermoviscous component. The determination of the two other relaxation frequencies and their compressibilities, as well as the compressibility contribution of the thermoviscous component are then decided by numerical minimization of the resulting difference to the desired power-law attenuation. %
For a large number of modeled relaxation mechanisms, such numerical optimization of the parameter fit turns very intricate.

Another approach is used in \cite{Kelly2009}, which is closely related to \cite{Schiessel1993, Schiessel95}. It demonstrates that hierarchical fractal ladder networks of springs and dashpots can lead to power-law acoustical attenuation in a low-frequency regime. This however requires a large number of degrees of freedom which makes parameter fits cumbersome. %

\section{The fractional Zener constitutive relation\label{sec:fractionalzener}}

The history of fractional derivatives in physics goes back to Abel's integral equation from 1826 \cite{Abel1826}, which turns out to correspond to the $1/2$-order derivative. %
%
Early viscoelasticity-related papers are \cite{Caputo1971, meshkov1971integral}, see also historical overviews in \cite{mainardi2012historical, Rossikhin2010B}. %
Inclusion of fractional derivatives in the viscoelastic stress-strain relationship is convenient for describing many materials where the response depends on the past history, see \eg{} \cite{Mainardi2010, mainardi2012historical, Podlubny1999chapter10_2} (for reflections on this from an acoustical point of view, see \cite{Treeby2010sectionIIB}). %
For a record on the most intuitive fractional generalizations of the conventional (non-fractional) viscoelastic models, see \eg{} the survey \cite{mainardi2011creeprelaxationviscosity}. 
In addition, the comprehensive reviews \cite{Rossikhin1997, Rossikhin2010}, summarize research on fractional calculus in dynamic problems of solid mechanics. %
Illustrations of the fractional derivative viscoelastic models commonly include the \emph{spring-pot} (or just \emph{pot}) element.

%
%
%
%
%
%
\subsection{Stress--strain relation}
%
%
%
A five-parameter fractional Zener model fractional generalization may be expressed as %
\begin{align}
	\sigma(t) +\tau_{\epsilon}^{\beta} \frac{\partial^{\beta}\sigma(t)}{\partial t^{\beta}}  = E_0 \left[\epsilon(t) +\tau_{\sigma}^{\alpha} \frac{\partial^{\alpha}\epsilon(t)}{\partial t^{\alpha}}\right],
	\label{Eq:gZener}
\end{align} 
where $t$ is the time, $\sigma(t)$ the stress, $\epsilon(t)$ the strain, $\taus$ and $\taue$ positive time constants, and $E_0$ the modulus. %
Here the nomenclature for the time-fractional derivatives follows \cite{Bagley83A}, however many authors display naming conventions where $\alpha$ and $\beta$ are interchanged. To be physically adequate, one requires $\alpha=\beta$ as further investigated in Section \ref{sec:parameter_restrictions} and Section \ref{sec:link_to_nachman}. %
The fractional Kelvin--Voigt constitutive relation may be regarded as a low-and intermediate frequency representation of the fractional Zener model \cite{Holm2011} which corresponds to $\taue \rightarrow 0$ in \eqref{Eq:gZener}. %

Other fractional stress-strain relations with either the same or more degrees of freedom may be used to describe material response, as stated in \eg{} \cite{Rossikhin2010}. One example is the 5-parameter approach described in ~\cite{Dinzart2009}. %
Such and other generalized models could equally well be applied in the wave equation derivations described in the following.

%
%
%
\subsection{Parameter value restrictions\label{sec:parameter_restrictions}}
%
%

Based on arguments from \cite{Bagley1986, Glockle1991}, a monotonically decreasing stress relaxation requires $\alpha=\beta$ in the stress-strain relation \eqref{Eq:gZener}. Table \ref{tab:constraints} lists thermodynamical of constraints on the \eqref{Eq:gZener} parameters. See also the related recent paper \cite{atanakovic2011thermodynamical}. %
	\begin{table}[htb] 
	\caption{\label{tab:constraints}Fractional Zener stress-strain model \eqref{Eq:gZener} parameter constraints, from \cite{Bagley1986}. }

	\centering
	\begin{tabular}{l@{\quad}   l@{\quad}  l@{\quad} l@{\quad} l@{\quad} l@{\quad} l@{\quad} l}
	\hline\hline
		\rule{0pt}{11pt}
		$E_0 \geq 0$, & $E_0\taus^\alpha>0$, & $\taus^\alpha \geq \taue^\beta$, &$\taue^\beta>0$, & $\alpha=\beta$\\
	\hline\hline
	\end{tabular}
\end{table}

We note that even though the rheological model \eqref{Eq:gZener} is not thermodynamically well-behaved for $\alpha\neq\beta$, the corresponding underlying mechanical model can actually be admissible. However this is only if the instantaneous wave is allowed to propagate at infinite speed \cite{rossikhin2001analysis}. 

The fractional Zener model with $\alpha=\beta$, which corresponds to the empirical Cole--Cole rheological relaxation spectrum formulation \cite{Cole1941}, was considered already in \cite{Caputo1971} alongside with experimental evidences. %

%
%
%
%
\subsection{Experimental evidences}\label{sec:experimental_evidences}
%
%
%
%

Parameter fits of experimental measurements to the fractional Zener model for biological materials include %
for %
brain \cite{Davis2006, Klatt2007, Kohandel2005, Sack2009}, %
human root dentin \cite{petrovic2005}, %
cranial bone \cite{liu2006}, %
liver \cite{Klatt2007}, 
arteries \cite{Craiem2008}, 
breast \cite{Coussot2009},
and hamstring muscle \cite{grahovac2010}. %
Non-biological materials are exemplified by 
metals \cite{Caputo1971}, %
doped corning glass \cite{Bagley83A}, %
rubber \cite{Bagley1986},
and polymers \cite{Coussot2009, Metzler2003, Pritz1996,Pritz2001, Pritz2003, sasso2011application}.
For an account of experimental fits to fractional calculus stress-strain models made up to the year of 2009, see Section 2 in \cite{Rossikhin2010}. %

The framework of viscoelasticity and acoustics in complex media has significant similarities with the framework of dielectrics and electromagnetic propagation. Not only do the wave equations have similarities in structure, but also the same constitutive relations connected to fractional derivative stress-strain modeling have experimentally been observed in the dielectrical properties of complex media. This is relevant for \eg{} ground-penetrating radar and medical diagnosis using ultrawideband or Terahertz electromagnetic waves. From an electromagnetic modeling point of view, the complex dielectric permittivity plays the same role as the compressibility (the complex compliance) does in acoustics and viscoelasticity.

\subsection{Maxwell--Wiechert representation}\label{sec:maxwellwiechert}

\begin{figure}[th!]
	\begin{center}
		\includegraphics[width=.7\columnwidth]{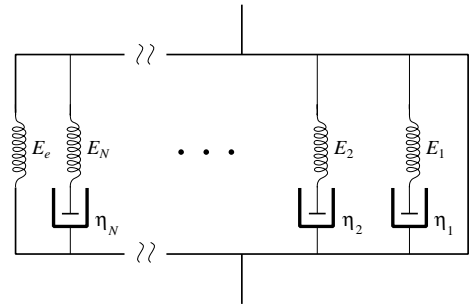}
	\end{center}
	\caption{Illustration of the Maxwell--Wiechert viscoelastic representation. Damper $i$ is denoted by $E_i$ and dashpot $i$ by $\eta_i$. \label{fig:MW} }
\end{figure}

%
Viscoelastic constitutive stress-strain models are generally possible to convert into a Prony series of Maxwell elements, that is a Maxwell--Wiechert description with springs and dashpots in parallel as illustrated in Fig.~\ref{fig:MW}. %

Already in \cite{mainardi1994fractional}, the fractional Zener model was actually interpreted as a relaxation distribution. %
Several other authors have published related results where the fractional derivative stress-strain models are expressed without using the exotic \emph{spring-pot} viscoelastic element. %
In \cite{Adolfsson2005}, a very large number of weighted Maxwell elements (Debye contributions) evenly distributed on a linear frequency scale are shown to give the same stress response as a fractional order viscoelastic model. %
Ref.~\cite{Chatterjee2005}, presented examples where viscoelastic damping due to several simultaneously decaying processes with closely spaced exponential decay rates are shown to induce a constitutive behavior involving fractional order derivatives. %
Furthermore, Machado and Galhano have shown that averaging over a large population of microelements, each having integer-order nature, gives global dynamics with both integer and fractional dynamics \cite{Machado2008}. %
We also note that the rheological fractional spring-pot element was interpreted in terms of weighted springs and dashpots in \cite{Papoulia2010}. %
The recent paper \cite{deOliveira2011} considers anomalous relaxation in dielectrics and interestingly provides relaxation distribution functions for non-Debye models flavors such as the Cole--Cole, Davidson--Cole, and Havriliak--Negami, out of which the Cole--Cole one is most tightly connected to the current work because it corresponds to $\alpha=\beta$ in the fractional Zener viscoelastic model. %

As further discussed in Section \ref{sec:link_to_nachman}, a Maxwell--Wiechert description of the fractional Zener model was actually implicitly verified also in \cite{Nasholm2011} where it was connected to the multiple relaxation framework of \cite{Nachman1990} via a distribution with fractal properties. %

\subsection{Physical background}
In \cite{bagley1991thermorheologically}, the author reflects on the surprisingly good fit of the fractional Zener model to measured data, suggesting that this hints at the existence of underlying governing principles. %
The proposed model is for internal energy loss at the molecular level of a viscoelastic polymer. %
A probability density to describe the motion of elevated energy states along the molecule or ``kinks'' was postulated. %
Such states behave similar to particles in potential wells which have to overcome barriers. By certain assumptions on probability density functions for both an energy transition and a distribution of barrier heights, it is shown that the probability density for an energy transition is fractal. %
The fractality leads to a relaxation function described by a power-law or a Mittag-Leffler function. This leads to a four parameter fractional Zener model linking stress and strain.

Another interpretation is due to Mainardi who justifies the four-parameter fractional Zener model from the thermoelastic equations when the temperature change due to diffusion and adiabatic strain change is a fractional derivative. The temperature change is thus a hidden variable in the stress-strain relationship \cite{mainardi1994fractional}.

A third point of view considers propagating waves. Attenuation of both compressional and shear waves is considered to be due to two mechanisms: Absorption
which is the conversion of energy into \eg{} heat, and scattering which is the reflection of energy in
all directions. Despite the different physical explanations, both mechanisms often seem to result in power law attenuation. 

In medical ultrasound below about 10 MHz, it is generally considered that absorption is the dominating mechanism \cite{bamber2005attenuationandabsorption}. It is likely that the above models are relevant for this regime. %
For elastography in tissue with propagating shear waves in the 10 and 1000 Hz range, recent experimental results indicate that attenuation due to multiple scattering can dominate \cite{chatelin2012measured, juge2012subvoxel}. %
A 1-D model for multiple scattering attenuation is the O'Doherty--Anstey model \cite{ODoherty71}, however there are no well established 2-D or 3-D model equivalents. Assuming a fractal distribution of reflection coefficients, the O'Doherty--Anstey model can explain power-law attenuation \cite{Garnier2010}. %

Connections between power-law attenuation, multiple scattering, and fractal geometry need to be further explored in real 2-D and 3-D media. 
%
%

\section{Deriving the fractional wave equation\label{sec:wave_equations}}
%
%
Below we demonstrate that causal wave equations can be constructed from the expressions for linearized conservation of momentum and the linearized conservation of mass combined with a fractional constitutive relation between stress and strain.
The approach outlined was applied already in \cite{Meidav1964} to derive a wave equation based on the non-fractional Zener model, which is often denoted as the \emph{standard linear solid}. %
%
%
\subsection{Conservation laws}
%
%
%
In the following, we consider an isotropic medium where the only non-negligible stiffness parameters are either the bulk modulus, $c_{11}$, or the shear modulus, $c_{44}$, \cite{Royer00}. Then the linearized conservation of mass corresponds to the strain being defined by
\begin{align}
	\epsilon(t) = \nabla u(x,t), \Fourierbackfourth \epsilon(\omega) = -i k\:u(k, \omega),
	\label{Eq:strain}
\end{align}
where $u$ is the displacement in the transverse (for compressional waves) or longitudinal (for shear waves) directions, $x$ is the 3-D spatial coordinate, and the symbol $\mathcal F$ denotes transformation into the spatio-temporal frequency domain where $\omega$ is the angular frequency and $k$ the wavenumber. %
This is valid under the assumption of infinitesimal strains and rotations which is adequate \eg{} in acoustical medical imaging.

The linearized conservation of momentum is expressed as
\begin{align}
	\nabla \sigma(t) = \rho_0 \frac{\partial^2u(x,t)}{\partial t^2} \Fourierbackfourth \ -i k \sigma(\omega) = \rho_0 (i \omega)^2 u(k,\omega),
	\label{Eq:Newton}
\end{align}
where $\rho_0$ is the steady-state mass density and $\sigma$ denotes the stress, which in this context corresponds to the negative of the pressure. %

%
%
%
\subsection{Generalized compressibility $\kappa(\omega)$}
%
%
%

The frequency-domain generalized compressibility is defined as the ratio between strain and stress: $\kappa(\omega) \triangleq \epsilon(\omega)/\sigma(\omega)$, therefore being related to the constitutive stress--strain relation. %
Combining this definition with the conservation laws \eqref{Eq:strain} and \eqref{Eq:Newton} gives 
\begin{align}
	\nabla^2u(x,t) = \dfrac{\dd^2}{\dd t^2} \left[\kappa(t) \underset{t}{*} u(x,t)\right]
	\Fourierbackfourth %
	\ k^2(\omega) = \omega^2\rho_0 \kappa(\omega) \label{eq:dispersion_relation}.
	%
\end{align}

Under circumstances where the linearized conservations of mass and momentum are valid, the wave equation is thus completely determined by the generalized compressibility. %

From \eqref{Eq:gZener}, the frequency-domain fractional Zener compressibility is obtained through the ratio $\epsilon(\omega)/\sigma(\omega)$:
\begin{align}
	\kappa_\text{Z}(\omega) & \triangleq %
		\kappaz \frac{1 + (\tau_{\epsilon}i \omega)^{\beta}}{1+ (\tau_{\sigma}i \omega)^{\alpha}}\notag\\
	&= %
		\kappaz -i\omega
	\kappa_0 \dfrac{ (i\omega)^{\alpha-1} - (\tau_\epsilon^\beta/\tau_\sigma^\alpha)(i\omega)^{\beta-1}}{ \tau_\sigma^{-\alpha} + (i\omega)^\alpha}
\label{Eq:fZener_compressibility}
\end{align}
The generalized compressibility $\kappa(\omega)$ as given above is (\eg{} in viscoelasticity) called complex compliance $J^*(\omega)=1/G^*(\omega)$, where $G^*(\omega)$ is the complex modulus. %
%

%
%
%
\subsection{The fractional Zener wave equation}
%
%
%

Insertion of the generalized compressibility \eqref{Eq:fZener_compressibility} into the dispersion relation \eqref{eq:dispersion_relation}, results in the time-domain fractional Zener wave equation \cite{Holm2011}
\begin{align}
	\nabla^2 u -\dfrac 1{c_0^2}\frac{\partial^2 u}{\partial t^2} + \taus^\alpha \dfrac{\partial^\alpha}{\partial t^\alpha}\nabla^2 u	- \dfrac {\taue^\beta}{c_0^2} \dfrac{\partial^{\beta+2} u}{\partial t^{\beta+2}} = 0.
	\label{Eq:wave_equation_zener}
\end{align}

\section{Properties of the fractional wave equation\label{sec:properties}}
Below we explore some properties of the fractional Zener wave equation, first by connecting it to a multiple relaxation wave model, then by studying the attenuation and the phase velocity as a function of frequency where 3 characteristic power-law regions are identified. The causality of the model is finally verified. %
More mathematics-oriented studies of fractional Zener-related wave equations can be found in \eg{} \cite{Atanackovic2002, Konjik2010, luchko2012fractional}. %

\subsection{Link to the NSW multiple relaxation framework\label{sec:link_to_nachman}}
%
Under the assumption of a continuum of relaxation mechanisms, %
the NSW model \cite{Nachman1990} is linked to fractional derivative modeling %
in \cite{Nasholm2011}, %
where it was shown that the wave equation corresponding to a certain distribution of relaxation contributions is identical to the fractional Zener wave equation. %
As resumed in the following, the associated compressibility contributions were shown to be distributed following a function related to the Mittag-Leffler function. %
The current section is conceptually tightly connected to Section \ref{sec:maxwellwiechert} where the fractional Zener constitutive model is represented as a set of springs and dashpots. 
%
%
%
%
%
\subsubsection{The NSW generalized compressibility\label{sec:nachman_generalized_compressibility}}
%
%

%
%
The NSW model of multiple discrete relaxation processes results in the generalized compressibility (which is equivalent to the rheological complex compliance $J^*(\omega)$) 
\begin{align}
	\kappao = \kappaz - i\omega \sum_{\nu=1}^N \dfrac{\kappan \taun}{1 + i\omega\taun},
	\label{eq:Nachman_kappa_omega}
\end{align}
where the mechanisms $\nu=1\ldots N$, have the relaxation times $\tau_1,\ldots,\tau_N$ and the compressibility contributions $\kappa_1, \ldots, \kappa_N$ \cite{Nachman1990}. %

Note that \eqref{eq:Nachman_kappa_omega} corresponds to a sum of $N$ weighted conventional Zener contributions, where each is given by \eqref{Eq:fZener_compressibility} with $\alpha=\beta=1$. %

Following \cite{Nasholm2011}, a representation of \eqref{eq:Nachman_kappa_omega} when considering a continuum of relaxation mechanisms distributed in the frequency band $\Omega \in[\Omega_1,\Omega_2]$ with the compressibility contributions described by the distribution $\kappa_\nu(\Omega)$ becomes
\begin{align}
	 \kappa_\text{N}(\omega) \triangleq \kappaz - i\omega\int_{\Omega_1}^{\Omega_2} \dfrac{ \kappan(\Omega) }{\Omega + i\omega}\, \dd \Omega.
	\label{eq:Nachman_kappa_omega_integral_zenertry}
\end{align}
Letting the limits of the integral go between $\Omega_1=0$ and $\Omega_2=\infty$, and instead incorporating any possible relaxation distribution bandwidth limitation into $\kappa_\nu(\Omega)$, the integral above is a Stieltjes transform. Applying the Laplace transform relation
\begin{align}
	\mathcal L^{-1}_\Omega\left\{\dfrac 1 {\Omega+i\omega}\right\}(t) = e^{-i\omega t},
\end{align}
by virtue of Fubini's theorem \cite{widder1971introductionchapter5_13} the generalized compressibility \eqref{eq:Nachman_kappa_omega_integral_zenertry} then becomes %
\begin{align}
	\kappa_\text{N}(\omega)
	&= \kappa_0 -i\omega \int_0^\infty \kappan(\Omega) \int_0^\infty e^{-\Omega t} e^{-i\omega t} \dd t\; \dd \Omega\notag\\
	\ &=  \kappa_0 -i\omega \mathcal F_t \Big\{H(t) \mathcal L_\Omega \left\{ \kappan(\Omega)  \right\}\!\! (t) \Big\}(\omega).
	\label{eq:nachman_fourier_laplace}
\end{align} 

Provided that the conservations of mass \eqref{Eq:strain} and momentum \eqref{Eq:Newton} are valid, and provided that the NSW generalized compressibility $\kappa_\text{N}(\omega)$ of \eqref{eq:nachman_fourier_laplace} 
is equal to the fractional Zener generalized compressibility $\kappa_\text{Z}(\omega)$ of \eqref{Eq:fZener_compressibility}, %
the dispersion relations from \eqref{eq:dispersion_relation} are also equal. Because the dispersion relation is a spatio-temporal Fourier representation of the wave equation, %
$\kappa_\text{N}(\omega) = \kappa_\text{Z}(\omega)$ thus implies that the NSW wave equation becomes equal to the fractional Zener wave equation \eqref{Eq:wave_equation_zener}. %
Direct comparison of $\kappa_\text{N}(\omega)$ in \eqref{eq:nachman_fourier_laplace} to $\kappa_\text{Z}(\omega)$ in \eqref{Eq:fZener_compressibility}, tells that they are equal in case the following is true:
\begin{align}
	\mathcal F_t &\Big\{H(t) \mathcal L_\Omega  \left\{ \kappan(\Omega)  \right\}\!\! (t) \Big\}(\omega)	
	=\notag\\
	&
	\kappa_0 \dfrac{ (i\omega)^{\alpha-1} - (\tau_\epsilon^\beta/\tau_\sigma^\alpha)(i\omega)^{\alpha-(\alpha-\beta+1)}}{ \tau_\sigma^{-\alpha} + (i\omega)^\alpha}.
	\label{eq:compressibilities_are_equal}
\end{align}
%
%

\subsubsection{The Cole--Cole equivalent $\alpha=\beta$ case}
%
First we choose to study the case $\alpha=\beta$ in a similar manner as in \cite{Nasholm2011}. Inverse Fourier transformation of both sides of \eqref{eq:compressibilities_are_equal}, then gives
\begin{align}
	H(t) \mathcal L_\Omega &\left\{ \kappan(\Omega)  \right\}\! (t)%
		=
	\kappa_0 (1-\tau_\epsilon^\alpha/\tau_\sigma^\alpha) \mathcal F_\omega^{-1} \left\{ \dfrac{ (i\omega)^{\alpha-1}}{ \tau_\sigma^{-\alpha} + (i\omega)^\alpha}\right\}\!(t)\notag\\
		&=
		\kappa_0 (1-\tau_\epsilon^\alpha/\tau_\sigma^\alpha) H(t) E_{\alpha,1} \left(-(t/\tau_\sigma)^\alpha \right),
	\label{eq:compressibilities_are_equal_alphaisbeta}
\end{align}
where $E_{a,b}( \cdot )$ is the Mittag-Leffler function (see Appendix \ref{section:ML_appendix}), and $H(t)$ is the Heaviside step function. The Fourier transform relation used in the last step above is given in \eqref{eq:mittagleffler_fourier_transform}. %
Moreover, using the inverse Laplace transform relation of \eqref{eq:mittagleffler_integral_relation}, Eq.~\eqref{eq:compressibilities_are_equal_alphaisbeta} hence gives
\begin{align}
	\kappan(\Omega) &=
	\kappa_0 (1-\tau_\epsilon^\alpha/\tau_\sigma^\alpha) f_{\alpha,1}\left(\Omega, \tau_\sigma^{-\alpha}\right)
	\notag\\
	& = \dfrac{1}{\pi} \dfrac{\kappaz(\taus^{\alpha}- \taue^\alpha)\Omega^{\alpha-1} \sin (\alpha\pi ) }{ (\taus\Omega)^{2\alpha} + 2(\taus\Omega)^\alpha \cos(\alpha\pi) + 1}
	\triangleq \kappa_{\nu\text{ML}}(\Omega) 
	\label{eq:distribution_for_alphaisbeta}
\end{align}
where $f_{\alpha,1}(\Omega,a)$ was inserted from \eqref{eq:f_alpha_beta_distribution}. %
Note that this link between the fractional Zener and the NSW models is valid also outside the small-attenuation regime $\Im\left\{k\right\} \ll \Re\left\{k\right\}$. 
Furthermore, it is worth emphasizing that the distribution function $\kappa_{\nu\text{ML}}(\Omega)$ has three distinct power-law regions where the transition is given by the product $\Omega\taus$:
\begin{align}
	\kappa_{\nu\text{ML}}(\Omega) \propto \left\{
		\begin{array}{ll}
			\displaystyle \Omega^{\alpha-1}, & \ \text{for } \Omega\taus \ll 1\\
			\displaystyle \Omega^{-1}, & \ \text{for } \Omega\taus \approx 1\\
			\displaystyle \Omega^{-\alpha-1},   & \  \text{for } 1 \ll \Omega\taus,
		\end{array}
		\right. 
	\label{eq:kappa_n_Omega_regimes}
\end{align}
We especially note that the high-frequency asymptote has fractal (self-similar) properties. Such fall-off also arises for Levy $\alpha$-stable distributions. %
This might reveal information on the underlying physics.

Keeping in mind that $\taue$ and $\taus$ may be regarded as break-frequencies, we note that fractional Zener time-parameters estimation don't really represent single relaxation times (frequencies) as for discrete Debye models, but rather break-times (frequencies) around which the model characteristics change. %

An inversion recipe to find the analogy of $\kappan(\Omega)$ given some attenuation law $\alpha_k(\omega)$ was presented in \cite{Vilensky2012}, by application of an approach tightly related to \cite{Pauly1971}. %
The small-attenuation assumption can at least for low frequencies often be reasonable for compressional wave propagation in biological tissue. By contrast, for shear-wave propagation in tissue, the attenuation is generally much more pronounced \cite{Szabo00}. %

%
%
%
%
\subsubsection{The $\alpha\neq \beta$ case} %
%
%
%
%

For the more general situation when $\alpha\neq \beta $, inverse Fourier transform on both sides of \eqref{eq:compressibilities_are_equal} instead gives%
\begin{align}
	H(t) &\mathcal L_\Omega \left\{ \kappan(\Omega)  \right\}\! (t)	=\notag\\
		&
	\kappa_0 \mathcal F_\omega^{-1} \left\{ \dfrac{ (i\omega)^{\alpha-1}}{ \tau_\sigma^{-\alpha} + (i\omega)^\alpha}\right\}\!(t) \notag\\
	&
	-\dfrac{\kappa_0 \tau_\epsilon^\beta}{\tau_\sigma^\alpha} \mathcal F_\omega^{-1} \left\{ \dfrac{ (i\omega)^{\alpha-(\alpha-\beta+1)}}{ \tau_\sigma^{-\alpha} + (i\omega)^\alpha}\right\}\!(t).
	\label{eq:compressibilities_are_equal_alphaisNOTbetafirst}
\end{align}
Subsequently exercising the inverse Fourier transforms gives
\begin{align}
	H&(t) \mathcal L_\Omega \left\{ \kappan(\Omega)  \right\}\! (t)	=\notag\\
		&
	H(t) \kappa_0 \Big[ E_{\alpha,1}\left(-(t/\taus)^\alpha\right)
	\notag\\
	&-(\tau_\epsilon^\beta/\tau_\sigma^\alpha)  t^{\alpha-\beta+1} E_{\alpha,\alpha-\beta+1}\left(-(t/\taus)^\alpha\right) \Big].
	\label{eq:compressibilities_are_equal_alphaisNOTbeta}
\end{align}
By recognizing in the equation above the Laplace transform relation \eqref{eq:mittagleffler_integral_relation} of the Appendix, the distribution which we choose to denote $\kappa_{\nu\text{ML}}'(\Omega)$ is identified:
\begin{align}
	\kappan(\Omega) &=
	\kappa_0 f_{\alpha,1}\left(\Omega, \tau_\sigma^{-\alpha}\right) 
		-\kappa_0(\tau_\epsilon^\beta/\tau_\sigma^\alpha) f_{\alpha,\alpha -\beta+1}\left(\Omega, \tau_\sigma^{-\alpha}\right)
	\notag\\
	=\ &  \dfrac{\kappa_0}{\pi \Omega} \cdot %
	\dfrac{1}{ (\taus\Omega)^{2\alpha} + 2(\taus\Omega)^\alpha \cos(\alpha\pi) + 1}%
	\cdot \notag\\
	&\cdot \Big[  (\tau_\sigma \Omega)^{\alpha} \sin (\alpha\pi )%
		%
		%
		%
		-(\tau_\epsilon\Omega)^{\beta}\sin(\beta\pi) \notag\\
		&\quad - (\tau_\sigma\Omega)^\alpha (\taue\Omega)^\beta \sin ((\beta-\alpha)\pi)%
		 \Big]
	\notag\\
	\triangleq\ & \kappa_{\nu\text{ML}}'(\Omega).
	\label{eq:distribution_for_alphaisNOTbeta}
\end{align}
The fractional Zener wave equation \eqref{Eq:wave_equation_zener} may at a first glance hence be contained within the NSW framework of multiple relaxation \cite{Nachman1990}, when assuming a continuum of relaxation mechanisms with the compressibility contribution as given by the distribution $\kappa_{\nu\text{ML}}'(\Omega)$ of \eqref{eq:distribution_for_alphaisNOTbeta}. %
We note that as a consequence of the $b<1$ criterion for the identity \eqref{eq:mittagleffler_integral_relation} to be valid, the step from \eqref{eq:compressibilities_are_equal_alphaisNOTbeta} to \eqref{eq:distribution_for_alphaisNOTbeta} is only correct for $\alpha \leq \beta$. %

On the other hand, for a continuous distribution of relaxation process contributions \eqref{eq:Nachman_kappa_omega_integral_zenertry} to be physically meaningful, the distribution $\kappa_\nu(\Omega)$ must be non-negative for all $\Omega$. A closer investigation of the distribution $\kappa_{\nu\text{ML}}'(\Omega)$ in \eqref{eq:distribution_for_alphaisNOTbeta} above reveals that no matter how the non-negative parameters are set, it cannot be positive for all $\Omega$. This is in accordance with the $\alpha = \beta$ thermodynamical restriction discussed in Section \ref{sec:parameter_restrictions}. %

This calls for a modification of the $\alpha \neq \beta$ version of the fractional Zener wave equation \eqref{Eq:wave_equation_zener} in order to make it thermodynamically admissible. Based on arguments presented in \cite{Pritz2003}, we suggest to start out from a modified, five-parameter form of the constitutive relation \eqref{Eq:gZener} which is equivalent to an ansatz analyzed in \cite{friedrich1992generalized} and reviewed in \cite{Rossikhin2010}:
\begin{align}
	\sigma(t) +\tau_{\epsilon}^{\beta} \frac{\partial^{\beta}\sigma(t)}{\partial t^{\beta}}  %
	= E_0 \left[\epsilon(t) %
	+ \tau_{\sigma}^{\alpha} \frac{\partial^{\alpha}\epsilon(t)}{\partial t^{\alpha}}%
	+ \tau_{\sigma}^{\beta} \frac{\partial^{\beta}\epsilon(t)}{\partial t^{\beta}}%
	\right],
	\label{eq:fractional_zener_modified}
\end{align} 
where we have $\alpha \leq \beta$. %
A related model has been applied to cell rheology \cite{djordjevic2003cell}. %
From the relation \eqref{eq:fractional_zener_modified} it is straightforward to construct a time-fractional wave equation using the methodology of Section \ref{sec:wave_equations} hence leading to
\begin{align}
	\nabla^2 u -\dfrac 1{c_0^2}\frac{\partial^2 u}{\partial t^2} %
	+ \taus^\alpha \dfrac{\partial^\alpha}{\partial t^\alpha}\nabla^2 u	%
	+ \taus^\beta \dfrac{\partial^\beta}{\partial t^\beta}\nabla^2 u	%
	- \dfrac {\taue^\beta}{c_0^2} \dfrac{\partial^{\beta+2} u}{\partial t^{\beta+2}} %
	= 0.
	\label{Eq:wave_equation_modified_zener}
\end{align}
We encourage researchers to execute the inverse transforms which yield the $\kappan(\Omega)$ NSW framework relaxation distribution corresponding to the above wave equation.

%
%
%
%
\subsubsection{Relation to the rheological relaxation time spectrum}
%
%
%
In viscoelasticity, a relaxation time spectrum, $\tilde H(\tau)$, related to $\kappa_\nu(\Omega)$ is commonly studied (see \eg{} \cite{Glockle1991} and references therein). It is related to the complex modulus through
\begin{align}
	G(t) = G_\infty + \int_{-\infty}^\infty  \tilde H(\tau) e^{-t/\tau} \dd \ln \tau.
\end{align}
It may be shown that for the fractional Zener model when setting $\Omega=\tau^{-1}$, the $\tau$-dependency of $\tilde H(\tau)$ differs by a factor $\tau$ from $\kappa_{\nu\text{ML}}(\Omega)$ of \eqref{eq:distribution_for_alphaisNOTbeta}. %
Figs.~5 and 6 of \cite{Glockle1991} illustrate that $\alpha=\beta$ gives symmetric $\tilde H(\tau)$, while $\alpha\neq\beta$ breaks the symmetry, most significantly far away from the peak region. %

%
%
%
\subsection{Attenuation and phase velocity}\label{sec:attenauation_and_phase_velocity}
%
%
%
The decomposition of the frequency-dependent wavenumber into its real and imaginary parts, 
gives the phase velocity $c_p(\omega) = \omega/\Re\left\{k(\omega)\right\}$ and attenuation $\alpha_k(\omega) = -\Im\left\{k(\omega)\right\}$. %
In general, the attenuation and the phase velocity are thus given by insertion of the generalized compressibility into the dispersion relation \eqref{eq:dispersion_relation} as
\begin{align}
\begin{array}{l}
	\alpha_k(\omega) = -\Im\left\{k\right\} = -\omega\sqrt{\rho_0}\Im\left\{\sqrt{\kappa(\omega)}\right\}	\quad\text{and}\\
	c_p(\omega) = \omega/\Re\left\{k\right\} = {\rho_0^{-1/2}}/\Re\left\{\sqrt{\kappa(\omega)}\right\}.
	\label{eq:atten_and_soundspeed_from_kappa}
	\end{array}
\end{align}

For the fractional Zener wave equation \eqref{Eq:wave_equation_zener} with $\alpha=\beta$, the attenuation expression \eqref{eq:atten_and_soundspeed_from_kappa} can be combined with the fractional Zener compressibility $\kappa_\text{Z}(\omega)$ of \eqref{Eq:fZener_compressibility} to get the attenuation. %
This results in three distinct frequency regimes of attenuation power-laws determined by the products $\taus\Omega$ and $\taue\Omega$ \cite{Nasholm2011}:
\begin{align}
	\alpha_k(\omega) \propto %
	\left\{
	\begin{array}{ll}
		\omega^{1+\alpha} & \text{low-frequencies,} \\
		\omega^{1-\alpha/2} & \text{intermediate frequencies,}\\
		\omega^{1-\alpha} & \text{high-frequencies.}
	\end{array}
	\right.
	\label{eq:freq_regions}
\end{align}
Below, the fractional Zener phase velocities and attenuations are further investigated numerically. %
The results from such calculations are compared to what is found by insertion of the distribution $\kappa_{\nu\text{ML}}(\Omega)$ of \eqref{eq:distribution_for_alphaisbeta} into the NSW generalized compressibility integral formula \eqref{eq:Nachman_kappa_omega_integral_zenertry}. This generalized compressibility is finally applied to \eqref{eq:atten_and_soundspeed_from_kappa}, from which $\alpha_k(\omega)$ and $c_p(\omega)$ are found. %

We use the latter calculation method to explore the effect of letting the continuum of relaxation mechanisms populate only a bounded frequency interval, rather than the entire $\Omega\in[0,\infty]$ region. %

\begin{figure}[th!]
	\begin{center}
		{\includegraphics[width=.81\columnwidth]{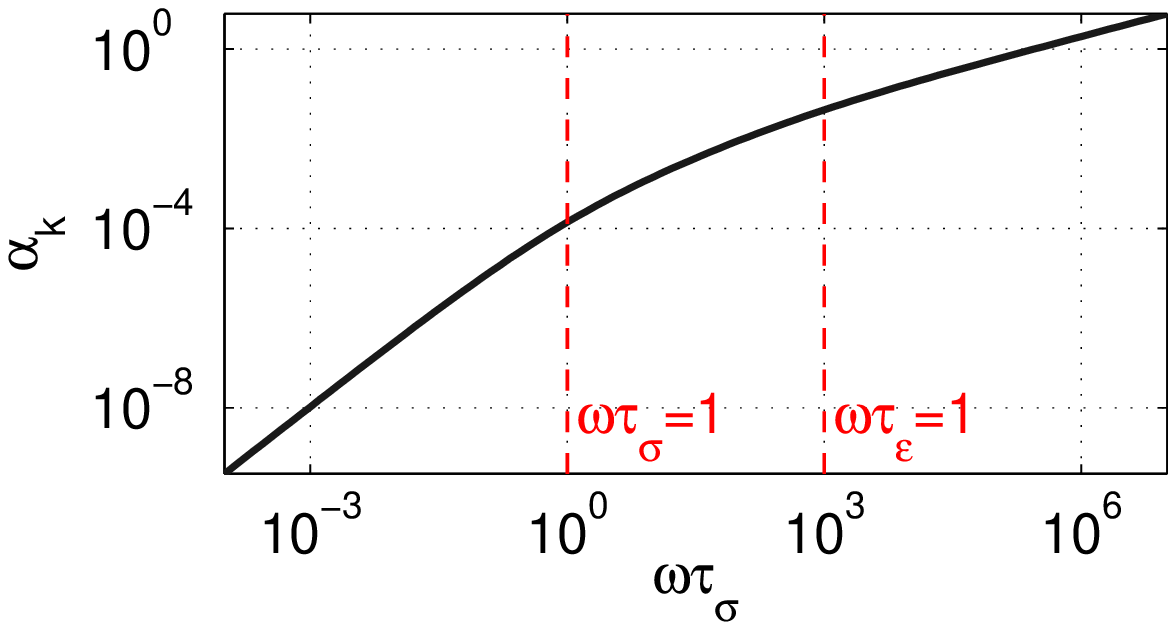}}\\
		\ \ \ {\includegraphics[width=.75\columnwidth]{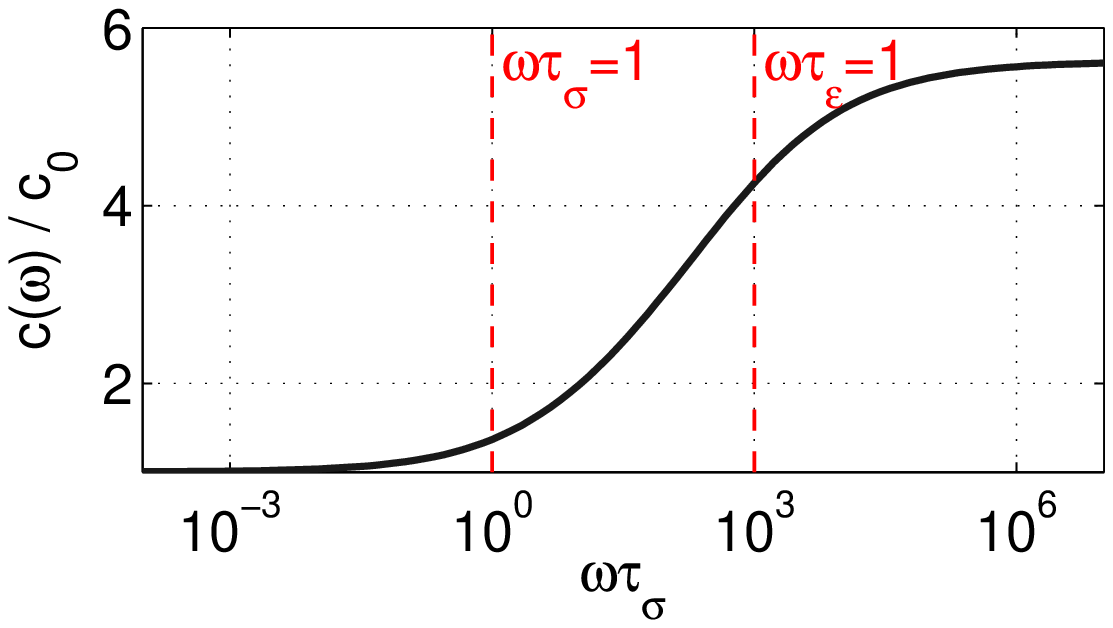}}\\
		{\includegraphics[width=.79\columnwidth]{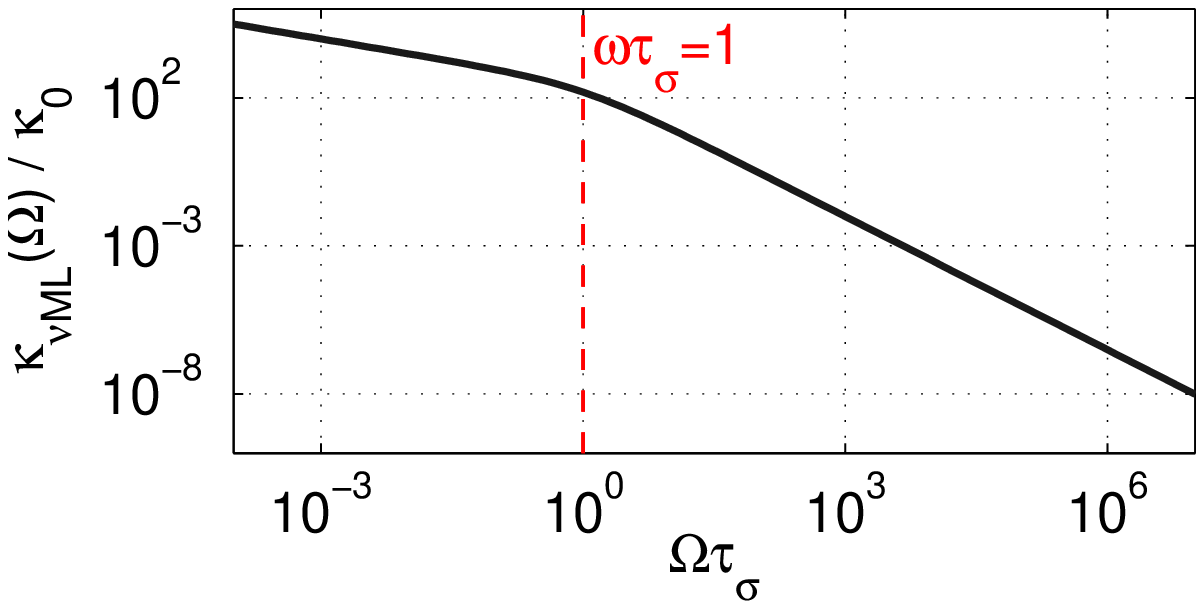}}
	\end{center} 
	\caption{\label{fig:curves}%
		Properties of the fractional Zener model exemplified for $\taus=1000\taue$ and $\alpha=\beta = 0.5$. %
		Top pane: frequency-dependent attenuation as a function of normalized wave frequency $\omega\cdot\taus$, where the three power-law regimes are distinguishable. %
		Middle pane: Frequency-dependent phase velocity as a function of normalized wave frequency $\omega\cdot\taus$. %
		Bottom pane: the corresponding normalized effective compressibilities $\kappa_{\nu\text{ML}}(\Omega)$ of the continuum of relaxation processes as a function of normalized relaxation frequency $\Omega\cdot\taus$. %
		} 
\end{figure}

Figure \ref{fig:curves} compares attenuation curves, the frequency-dependent phase velocity, and the distribution $\kappa_{\nu\text{ML}}(\Omega)$. %
Note that for many applications, the ratio $\taus/\taue$ is only slightly larger than one. This implies that the intermediate regime becomes negligible. %
For attenuation in seawater \cite{Ainslie1998} and air \cite{Bass1995}, one usually considers only three discrete relaxation contributions each with $\alpha=1$, which results in the familiar $\alpha_k \propto \omega^2$ for low frequencies and constant attenuation for high frequencies. %

%
%
%

In the figures we observe that the parameter $\taus$ may be regarded as related to break frequencies, where the distribution of relaxation frequencies crosses over between different power-law regimes of $\kappa_\nu(\Omega)$. %
Notably the low- and high-frequency asymptotes of $\kappa_\nu(\Omega)$, which both have fractal properties, are also well visible. %
The break frequencies also correspond to where the attenuation crosses over between different power-law regimes. %

%
%
%
%
\subsection{Causality and finite phase speed}
%
%
%
%

In particular, we observe that according to the NSW paper \cite{Nachman1990} any physical mechanism that fits into the multiple relaxation framework corresponds to finite phase speed, non-negative attenuation, and causal response at all wave frequencies. %
This parallels the conversion of stress-strain models into the Maxwell--Wiechert representation. %
The NSW model causality is also verified because it complies with the Kramers--Kronig causality relations. See \cite{Waters2005} for an acoustics-oriented treatment of these relations. %

In Section \ref{sec:nachman_generalized_compressibility}, we re-wrote the continuous relaxation process distribution \eqref{eq:Nachman_kappa_omega_integral_zenertry}, which is a generalization of the NSW discrete sum \eqref{eq:Nachman_kappa_omega}, %
into expression \eqref{eq:nachman_fourier_laplace}: $\kappa_\text{N}(\omega) = \kappa_0 -i\omega \mathcal F_t \Big\{H(t) \mathcal L_\Omega \left\{ \kappan(\Omega)  \right\}\!\! (t) \Big\}(\omega)$. %
Studying this expression sparks the conclusion that any distribution of relaxation contributions $\kappan(\Omega)$ for which the successive Laplace and Fourier integrals of \eqref{eq:nachman_fourier_laplace} exist, the corresponding wave equation gives finite phase speed, non-negative attenuation, and causal solutions for all wave frequencies. %
Moreover, it is worth emphasizing that some models are causal but can imply unbounded phase speeds. For example the fractional Kelvin--Voigt wave equation the unbounded phase speed at infinite frequencies \cite{Caputo1967, Wismer06, Holm2010}. %
The causality properties of several acoustical attenuation models are investigated in \cite{kowar2012attenuation}. %
Regarding restrictions on the attenuation power-law exponents, \cite{Weaver1981} argues that causality is maintained only if the attenuation has a slower than linear rise with frequency in the high-frequency limit. This requirement is met both for the fractional Zener wave attenuation and the NSW attenuation. %
See also \cite{seredynska2010relaxationdispersion} for a related study.

%
%
%
%
\section{Discussion and concluding remarks}
%
%
%
%

Within the framework of elastic waves, the current work surveys connections between concepts of power-law and multiple relaxation attenuation, causality, and fractional derivative differential equations. %

The fractional Zener elastic wave equation model fits attenuation measurements well and is characterized by a small number of parameters. The NSW model \cite{Nachman1990} is widely acknowledged in acoustical modeling and does not comprise fractional derivatives. %
Here we have analyzed how the fractional Zener \eqref{Eq:gZener} and NSW \eqref{eq:Nachman_kappa_omega_integral_zenertry} wave equation models can be unified under the assumption of a continuous distribution of relaxation mechanisms \eqref{eq:distribution_for_alphaisbeta} which has fractal properties. %

Because NSW-compatible wave equations are causal as well as consistent with non-negative attenuation and finite phase speed, we prefer to consider any such model as eligible form an physically intuitive point of view. %
Nevertheless, it is still unclear to the acoustics community what are the underlying physical mechanisms which interplay in complex media to result in power-law attenuation of elastic waves. %

The characteristics of viscoelasticity and acoustics of complex media, share many similarities with what is encountered for dielectrical properties of complex media. %
We here point out that there are several theories within this field on how to explain \eg{} Cole--Cole behavior by medium disordering, scaling, and geometry as well as more probabilistic approaches. See \cite{weron2000probabilistic, nigmatullin2005theoryofdielectric, stanislavsky2007stochasticnature, feldman2012dielectric} and the references therein. Maybe the search for first principles explanations for the fractional behavior of complex elastic media can be inspired by such findings. %

Furthermore, we note that relaxation processes in nuclear magnetic resonance (NMR) as often described by the so called Bloch equations actually also can give non-Debye appearance, see \eg{} \cite{bhalekar2012generalizedfractional} and the references therein. %
In addition, developments related to dispersion and attenuation of elastic waves have many traits in common with the mathematical descriptions of luminescence decay, see \eg{} \cite{berberiansantos2008}. %

We aim to encourage the acoustical community to more frequently adopt fractional calculus descriptions for wave modeling in complex media.  
Hopefully we can also stimulate both mathematics-oriented and other researchers to spark further progress within fractional modeling of elastic waves by contributing to advance in e.g.\ model enhancements, existence and Green's function considerations, as well as in analytical and numerical investigations.
%
We call for further exploration of connections between fractional dynamics, the surprisingly prevalent power-law patterns of nature, and the micromechanical structure of complex materials.

\section{Appendices}
\subsection{Appendix: Mittag-Leffler function properties}
{\label{section:ML_appendix}
%
%
%
\subsubsection{Definition and Fourier Transform Relation}
The one-parameter Mittag-Leffler function was introduced in \cite{Mittag-Leffler1903}. A two-parameter analogy was presented in \cite{Wiman1905}, and may be written as
\begin{align}
	\MittagLefflerAlphaBetaT \triangleq \sum_{n=0}^\infty \dfrac{t^n}{\Gamma(a n+ b )},
	\label{eq:mittagleffler_definition}
\end{align}
where $\Gamma$ is the Euler Gamma function and the parameters are commonly restricted to $\{a, b \} \in \C,\ \Re{\{a, b \}}>0$, and $t\in\C$. %
See~\cite{Haubold2011} for a comprehensive review of Mittag-Leffler function properties. 

A useful Fourier transform pair involving the Mittag-Leffler function is \cite{Podlubny1999chapter1-2}
\begin{align} 
	\Fouriertransform{H(t)\: t^{ b -1}\MittagLeffler{ a }{ b }{-A t^ a }}(\omega) = & \dfrac{(i\omega)^{ a - b }}{A+ (i\omega)^ a }.
	\label{eq:mittagleffler_fourier_transform}
\end{align}
%

\subsubsection{\label{section:ML_appendix_integral_representation}Laplace Transform Integral Representation}
%
%
%
%
The function $t^{b-1}\MittagLeffler{ a }{ b }{{-A}t^a}$ may for $0< a \leq b < 1$ be written on an integral form \cite{Djrbashian1966,Djrbashian1993chapter1}:
\begin{align}
	t^{ b -1}\MittagLeffler{ a }{ b }{-A t^ a } = \int_0^\infty e^{-\Omega t} f_{ a , b }(\Omega, A)\: \dd \Omega,
	\label{eq:mittagleffler_integral_relation}
\end{align}
where 
\begin{align}
	f_{ a , b }(\Omega,A) = \dfrac{\Omega^{ a - b }}{\pi} \dfrac{A \sin [( b - a )\pi ] + \Omega^ a  \sin( b \pi)}{ \Omega^{2 a } + 2 A \Omega^ a  \cos( a \pi) + A^2}.
		\label{eq:f_alpha_beta_distribution}
\end{align}
Careful reading of Appendix E in \cite{Mainardi2010} reveals that the above above functions $f_{a,b}(\Omega,A)$ may be denoted \emph{spectral functions}, which are non-negative. %
%

\section*{Acknowledgements}

The authors would like to thank the FCAA editor and reviewers for recommending us to submit this Survey paper. %
This paper has been partially supported by the ``High Resolution Imaging and Beamforming'' project of the Norwegian Research Council.

\bibliographystyle{jasanum}

\begin{thebibliography}{100}
\newcommand{\enquote}[1]{``#1''}
\expandafter\ifx\csname url\endcsname\relax
  \def\url#1{\texttt{#1}}\fi
\expandafter\ifx\csname urlprefix\endcsname\relax\def\urlprefix{URL }\fi
\providecommand{\bibinfo}[2]{#2}
\providecommand{\noopsort}[1]{}
\providecommand{\switchargs}[2]{#2#1}

\bibitem{Nasholm2012China}
\bibinfo{author}{S.~P. N\"asholm} and \bibinfo{author}{S.~Holm},
  \enquote{\bibinfo{title}{A fractional acoustic wave equation from multiple
  relaxation loss and conservation laws}}, \bibinfo{journal}{Proc. 5th Int.
  Workshop on Fractional Differentiation and its Applications}
  (\bibinfo{year}{2012}).

\bibitem{Holm2011}
\bibinfo{author}{S.~Holm} and \bibinfo{author}{S.~P. N\"asholm},
  \enquote{\bibinfo{title}{A causal and fractional all-frequency wave equation
  for lossy media}}, \bibinfo{journal}{J.\ Acoust.\ Soc.\ Am.}
  \textbf{\bibinfo{volume}{130}}, \bibinfo{pages}{2195--2202}
  (\bibinfo{year}{2011}).

\bibitem{Nasholm2011}
\bibinfo{author}{S.~P. N\"asholm} and \bibinfo{author}{S.~Holm},
  \enquote{\bibinfo{title}{Linking multiple relaxation, power-law attenuation,
  and fractional wave equations}}, \bibinfo{journal}{J.\ Acoust.\ Soc.\ Am.}
  \textbf{\bibinfo{volume}{130}}, \bibinfo{pages}{3038--3045}
  (\bibinfo{year}{2011}).

\bibitem{Nachman1990}
\bibinfo{author}{A.~I. Nachman}, \bibinfo{author}{J.~F. Smith~III}, and
  \bibinfo{author}{R.~C. Waag}, \enquote{\bibinfo{title}{An equation for
  acoustic propagation in inhomogeneous media with relaxation losses}},
  \bibinfo{journal}{J.\ Acoust.\ Soc.\ Am.} \textbf{\bibinfo{volume}{88}},
  \bibinfo{pages}{1584--1595} (\bibinfo{year}{1990}).

\bibitem{Duck1990}
\bibinfo{author}{F.~A. Duck}, \emph{\bibinfo{title}{Physical properties of
  tissue}} (\bibinfo{publisher}{Academic Press}) (\bibinfo{year}{1990}).

\bibitem{Szabo00}
\bibinfo{author}{T.~L. Szabo} and \bibinfo{author}{J.~Wu},
  \enquote{\bibinfo{title}{A model for longitudinal and shear wave propagation
  in viscoelastic media}}, \bibinfo{journal}{J.\ Acoust.\ Soc.\ Am.}
  \textbf{\bibinfo{volume}{107}}, \bibinfo{pages}{2437--2446}
  (\bibinfo{year}{2000}).

\bibitem{Holm2010}
\bibinfo{author}{S.~Holm} and \bibinfo{author}{R.~Sinkus},
  \enquote{\bibinfo{title}{{A unifying fractional wave equation for
  compressional and shear waves}}}, \bibinfo{journal}{J.\ Acoust.\ Soc.\ Am.}
  \textbf{\bibinfo{volume}{127}}, \bibinfo{pages}{542--548}
  (\bibinfo{year}{2010}).

\bibitem{beard2011biomedical}
\bibinfo{author}{P.~Beard}, \enquote{\bibinfo{title}{Biomedical photoacoustic
  imaging}}, \bibinfo{journal}{Interface Focus} \textbf{\bibinfo{volume}{1}},
  \bibinfo{pages}{602--631} (\bibinfo{year}{2011}).

\bibitem{kowar2012attenuation}
\bibinfo{author}{R.~Kowar} and \bibinfo{author}{O.~Scherzer},
  \enquote{\bibinfo{title}{Attenuation models in photoacoustics}}, in
  \emph{\bibinfo{booktitle}{Mathematical Modeling in Biomedical Imaging II}},
  edited by \bibinfo{editor}{H.~Ammari}, volume \bibinfo{volume}{2035} of
  \emph{\bibinfo{series}{Lecture Notes in Mathematics}},
  \bibinfo{pages}{85--130} (\bibinfo{publisher}{Springer Berlin / Heidelberg})
  (\bibinfo{year}{2012}).

\bibitem{roitner2012experimental}
\bibinfo{author}{H.~Roitner}, \bibinfo{author}{J.~Bauer-Marschallinger},
  \bibinfo{author}{T.~Berer}, and \bibinfo{author}{P.~Burgholzer},
  \enquote{\bibinfo{title}{Experimental evaluation of time domain models for
  ultrasound attenuation losses in photoacoustic imaging}},
  \bibinfo{journal}{J.\ Acoust.\ Soc.\ Am.} \textbf{\bibinfo{volume}{131}},
  \bibinfo{pages}{3763--3774} (\bibinfo{year}{2012}).

\bibitem{treeby2010reconstruction}
\bibinfo{author}{B.~E. Treeby}, \bibinfo{author}{E.~Z. Zhang}, and
  \bibinfo{author}{B.~T. Cox}, \enquote{\bibinfo{title}{Photoacoustic
  tomography in absorbing acoustic media using time reversal}},
  \bibinfo{journal}{Inverse Probl.} \textbf{\bibinfo{volume}{26}},
  \bibinfo{pages}{115003} (\bibinfo{year}{2010}).

\bibitem{2012ehmanreview}
\bibinfo{author}{R.~L. Ehman}, \bibinfo{author}{K.~J. Glaser}, and
  \bibinfo{author}{A.~Manduca}, \enquote{\bibinfo{title}{Review of {MR}
  elastography applications and recent developments}}, \bibinfo{journal}{J.
  Magn. Reson.} \textbf{\bibinfo{volume}{36}}, \bibinfo{pages}{757--774}
  (\bibinfo{year}{2012}).

\bibitem{muthupillai1995science}
\bibinfo{author}{R.~Muthupillai}, \bibinfo{author}{D.~J. Lomas},
  \bibinfo{author}{P.~J. Rossman}, \bibinfo{author}{J.~F. Greenleaf},
  \bibinfo{author}{A.~Manduca}, and \bibinfo{author}{R.~L. Ehman},
  \enquote{\bibinfo{title}{Magnetic resonance elastography by direct
  visualization of propagating acoustic strain waves}},
  \bibinfo{journal}{Science} \textbf{\bibinfo{volume}{269}},
  \bibinfo{pages}{1854--1857} (\bibinfo{year}{1995}).

\bibitem{papazoglou2012multifrequency}
\bibinfo{author}{S.~Papazoglou}, \bibinfo{author}{S.~Hirsch},
  \bibinfo{author}{J.~Braun}, and \bibinfo{author}{I.~Sack},
  \enquote{\bibinfo{title}{Multifrequency inversion in magnetic resonance
  elastography}}, \bibinfo{journal}{Phys. Med. Biol.}
  \textbf{\bibinfo{volume}{57}}, \bibinfo{pages}{2329--2346}
  (\bibinfo{year}{2012}).

\bibitem{Sinkus2000}
\bibinfo{author}{R.~Sinkus}, \bibinfo{author}{J.~Lorenzen},
  \bibinfo{author}{D.~Schrader}, \bibinfo{author}{M.~Lorenzen},
  \bibinfo{author}{M.~Dargatz}, and \bibinfo{author}{D.~Holz},
  \enquote{\bibinfo{title}{High-resolution tensor {MR} elastography for breast
  tumour detection}}, \bibinfo{journal}{Phys. Med. Biol.}
  \textbf{\bibinfo{volume}{45}}, \bibinfo{pages}{1649--1664}
  (\bibinfo{year}{2000}).

\bibitem{Sinkus2007}
\bibinfo{author}{R.~Sinkus}, \bibinfo{author}{K.~Siegmann},
  \bibinfo{author}{T.~Xydeas}, \bibinfo{author}{M.~Tanter},
  \bibinfo{author}{C.~Claussen}, and \bibinfo{author}{M.~Fink},
  \enquote{\bibinfo{title}{{{MR} elastography of breast lesions: Understanding
  the solid/liquid duality can improve the specificity of contrast-enhanced
  {MR} mammography}}}, \bibinfo{journal}{Magn. Res. in Med.}
  \textbf{\bibinfo{volume}{58}}, \bibinfo{pages}{1135--1144}
  (\bibinfo{year}{2007}).

\bibitem{sinkus2012review}
\bibinfo{author}{R.~Sinkus}, \bibinfo{author}{J.-L. Daire},
  \bibinfo{author}{V.~Vilgrain}, and \bibinfo{author}{B.~E. Van~Beers},
  \enquote{\bibinfo{title}{Elasticity imaging via {MRI}: Basics, overcoming the
  waveguide limit, and clinical liver results}}, \bibinfo{journal}{Curr. Med.
  Imaging Rev.} \textbf{\bibinfo{volume}{8}}, \bibinfo{pages}{56--63}
  (\bibinfo{year}{2012}).

\bibitem{yasar2012wideband}
\bibinfo{author}{T.~K. Yasar}, \bibinfo{author}{T.~J. Royston}, and
  \bibinfo{author}{R.~L. Magin}, \enquote{\bibinfo{title}{Wideband {MR}
  elastography for viscoelasticity model identification}},
  \bibinfo{journal}{Magnet. Reson. Med}  (\bibinfo{year}{2012}),
  \bibinfo{note}{online Version of Record published before inclusion in an
  issue}.

\bibitem{barry2012shearwavesispersion}
\bibinfo{author}{C.~T. Barry}, \bibinfo{author}{B.~Mills},
  \bibinfo{author}{Z.~Hah}, \bibinfo{author}{R.~A. Mooney},
  \bibinfo{author}{C.~K. Ryan}, \bibinfo{author}{D.~J. Rubens}, and
  \bibinfo{author}{K.~J. Parker}, \enquote{\bibinfo{title}{Shear wave
  dispersion measures liver steatosis}}, \bibinfo{journal}{Ultrasound Med.
  Biol.} \textbf{\bibinfo{volume}{38}}, \bibinfo{pages}{175--182}
  (\bibinfo{year}{2012}).

\bibitem{bercoff2004supersonic}
\bibinfo{author}{J.~Bercoff}, \bibinfo{author}{M.~Tanter}, and
  \bibinfo{author}{M.~Fink}, \enquote{\bibinfo{title}{Supersonic shear imaging:
  a new technique for soft tissue elasticity mapping}}, \bibinfo{journal}{IEEE
  Trans.\ Ultrason.\ Ferroelectr.,\ Freq.\ Control}
  \textbf{\bibinfo{volume}{51}}, \bibinfo{pages}{396--409}
  (\bibinfo{year}{2004}).

\bibitem{Chen2004}
\bibinfo{author}{S.~Chen}, \bibinfo{author}{M.~Fatemi}, and
  \bibinfo{author}{J.~F. Greenleaf}, \enquote{\bibinfo{title}{Quantifying
  elasticity and viscosity from measurement of shear wave speed dispersion}},
  \bibinfo{journal}{J.\ Acoust.\ Soc.\ Am.} \textbf{\bibinfo{volume}{115}},
  \bibinfo{pages}{2781--2785} (\bibinfo{year}{2004}).

\bibitem{palmeri2011acoustic}
\bibinfo{author}{M.~L. Palmeri} and \bibinfo{author}{K.~R. Nightingale},
  \enquote{\bibinfo{title}{Acoustic radiation force-based elasticity imaging
  methods}}, \bibinfo{journal}{Interface Focus} \textbf{\bibinfo{volume}{1}},
  \bibinfo{pages}{553--564} (\bibinfo{year}{2011}).

\bibitem{Wismer1995}
\bibinfo{author}{M.~G. Wismer} and \bibinfo{author}{R.~Ludwig},
  \enquote{\bibinfo{title}{An explicit numerical time domain formulation to
  simulate pulsed pressure waves in viscous fluids exhibiting arbitrary
  frequency power law attenuation}}, \bibinfo{journal}{IEEE Trans.\ Ultrason.\
  Ferroelectr.,\ Freq.\ Control} \textbf{\bibinfo{volume}{42}},
  \bibinfo{pages}{1040--1049} (\bibinfo{year}{1995}).

\bibitem{Liebler2004}
\bibinfo{author}{M.~Liebler}, \bibinfo{author}{S.~Ginter},
  \bibinfo{author}{T.~Dreyer}, and \bibinfo{author}{R.~E. Riedlinger},
  \enquote{\bibinfo{title}{Full wave modeling of therapeutic ultrasound:
  {E}fficient time-domain implementation of the frequency power-law
  attenuation}}, \bibinfo{journal}{J.\ Acoust.\ Soc.\ Am.}
  \textbf{\bibinfo{volume}{116}}, \bibinfo{pages}{2742--2750}
  (\bibinfo{year}{2004}).

\bibitem{Wismer06}
\bibinfo{author}{M.~G. Wismer}, \enquote{\bibinfo{title}{Finite element
  analysis of broadband acoustic pulses through inhomogenous media with power
  law attenuation}}, \bibinfo{journal}{J.\ Acoust.\ Soc.\ Am.}
  \textbf{\bibinfo{volume}{120}}, \bibinfo{pages}{3493--3502}
  (\bibinfo{year}{2006}).

\bibitem{Caputo2011}
\bibinfo{author}{M.~Caputo}, \bibinfo{author}{J.~M. Carcione}, and
  \bibinfo{author}{F.~Cavallini}, \enquote{\bibinfo{title}{Wave simulation in
  biologic media based on the {K}elvin--{V}oigt fractional-derivative
  stress--strain relation}}, \bibinfo{journal}{Ultrasound Med. Biol.}
  \textbf{\bibinfo{volume}{37}}, \bibinfo{pages}{996--1004}
  (\bibinfo{year}{2011}).

\bibitem{Prieur2011}
\bibinfo{author}{F.~Prieur} and \bibinfo{author}{S.~Holm},
  \enquote{\bibinfo{title}{Nonlinear acoustic wave equations with fractional
  loss operators}}, \bibinfo{journal}{J.\ Acoust.\ Soc.\ Am.}
  \textbf{\bibinfo{volume}{130}}, \bibinfo{pages}{1125--1132}
  (\bibinfo{year}{2011}).

\bibitem{Prieur2012}
\bibinfo{author}{F.~Prieur}, \bibinfo{author}{G.~Vilenskiy}, and
  \bibinfo{author}{S.~Holm}, \enquote{\bibinfo{title}{A more fundamental
  approach to the derivation of nonlinear acoustic wave equations with
  fractional loss operators}}, \bibinfo{journal}{J.\ Acoust.\ Soc.\ Am.}
  \textbf{\bibinfo{volume}{132}}, \bibinfo{pages}{2169--2172}
  (\bibinfo{year}{2012}).

\bibitem{treeby2012modeling}
\bibinfo{author}{B.~E. Treeby}, \bibinfo{author}{J.~Jaros},
  \bibinfo{author}{A.~P. Rendell}, and \bibinfo{author}{B.~T. Cox},
  \enquote{\bibinfo{title}{Modeling nonlinear ultrasound propagation in
  heterogeneous media with power law absorption using a $k$-space
  pseudospectral method}}, \bibinfo{journal}{J.\ Acoust.\ Soc.\ Am.}
  \textbf{\bibinfo{volume}{131}}, \bibinfo{pages}{4324--4336}
  (\bibinfo{year}{2012}).

\bibitem{Chen03}
\bibinfo{author}{W.~Chen} and \bibinfo{author}{S.~Holm},
  \enquote{\bibinfo{title}{Modified {S}zabo's wave equation models for lossy
  media obeying frequency power law}}, \bibinfo{journal}{J.\ Acoust.\ Soc.\
  Am.} \textbf{\bibinfo{volume}{114}}, \bibinfo{pages}{2570--2574}
  (\bibinfo{year}{2003}).

\bibitem{Treeby2010}
\bibinfo{author}{B.~E. Treeby} and \bibinfo{author}{B.~T. Cox},
  \enquote{\bibinfo{title}{{Modeling power law absorption and dispersion for
  acoustic propagation using the fractional Laplacian}}}, \bibinfo{journal}{J.\
  Acoust.\ Soc.\ Am.} \textbf{\bibinfo{volume}{127}},
  \bibinfo{pages}{2741--2748} (\bibinfo{year}{2010}).

\bibitem{Carcione2010}
\bibinfo{author}{J.~M. Carcione}, \enquote{\bibinfo{title}{A generalization of
  the {F}ourier pseudospectral method}}, \bibinfo{journal}{Geophysics}
  \textbf{\bibinfo{volume}{75}}, \bibinfo{pages}{A53--A56}
  (\bibinfo{year}{2010}).

\bibitem{Tabei2003}
\bibinfo{author}{M.~Tabei}, \bibinfo{author}{T.~D. Mast}, and
  \bibinfo{author}{R.~C. Waag}, \enquote{\bibinfo{title}{Simulation of
  ultrasonic focus aberration and correction through human tissue}},
  \bibinfo{journal}{J.\ Acoust.\ Soc.\ Am.} \textbf{\bibinfo{volume}{113}},
  \bibinfo{pages}{1166--1176} (\bibinfo{year}{2003}).

\bibitem{Yang2005}
\bibinfo{author}{X.~Yang} and \bibinfo{author}{R.~O. Cleveland},
  \enquote{\bibinfo{title}{Time domain simulation of nonlinear acoustic beams
  generated by rectangular pistons with application to harmonic imaging}},
  \bibinfo{journal}{J.\ Acoust.\ Soc.\ Am.} \textbf{\bibinfo{volume}{117}},
  \bibinfo{pages}{113--123} (\bibinfo{year}{2005}).

\bibitem{Kelly2009}
\bibinfo{author}{J.~F. Kelly} and \bibinfo{author}{R.~J. McGough},
  \enquote{\bibinfo{title}{Fractal ladder models and power law wave
  equations}}, \bibinfo{journal}{J.\ Acoust.\ Soc.\ Am.}
  \textbf{\bibinfo{volume}{126}}, \bibinfo{pages}{2072--2081}
  (\bibinfo{year}{2009}).

\bibitem{Schiessel1993}
\bibinfo{author}{H.~Schiessel} and \bibinfo{author}{A.~Blumen},
  \enquote{\bibinfo{title}{Hierarchical analogues to fractional relaxation
  equations}}, \bibinfo{journal}{J. Phys. A} \textbf{\bibinfo{volume}{26}},
  \bibinfo{pages}{5057--5069} (\bibinfo{year}{1993}).

\bibitem{Schiessel95}
\bibinfo{author}{H.~Schiessel} and \bibinfo{author}{A.~Blumen},
  \enquote{\bibinfo{title}{{Mesoscopic Pictures of Sol-Gel Transition: Ladder
  Models and Fractal Networks}}}, \bibinfo{journal}{Macromolecules}
  \textbf{\bibinfo{volume}{28}}, \bibinfo{pages}{4013--4019}
  (\bibinfo{year}{1995}).

\bibitem{Abel1826}
\bibinfo{author}{N.~H. Abel}, \enquote{\bibinfo{title}{Aufl{\"o}sung einer
  mechanischen {A}ufgabe ({R}esolution of a mechanical problem)}},
  \bibinfo{journal}{J.~{R}eine.~{A}ngew.~{M}ath.} \bibinfo{pages}{153--157}
  (\bibinfo{year}{1826}).

\bibitem{Caputo1971}
\bibinfo{author}{M.~Caputo} and \bibinfo{author}{F.~Mainardi},
  \enquote{\bibinfo{title}{A new dissipation model based on memory mechanism}},
  \bibinfo{journal}{Pure Appl. Geophys.} \textbf{\bibinfo{volume}{91}},
  \bibinfo{pages}{134--147} (\bibinfo{year}{1971}).

\bibitem{meshkov1971integral}
\bibinfo{author}{S.~I. Meshkov}, \bibinfo{author}{G.~N. Pachevskaya},
  \bibinfo{author}{V.~S. Postnikov}, and \bibinfo{author}{U.~A. Rossikhin},
  \enquote{\bibinfo{title}{Integral representations of
  ${\owns}_\gamma$--functions and their application to problems in linear
  viscoelasticity}}, \bibinfo{journal}{Int. J. Eng. Sci.}
  \textbf{\bibinfo{volume}{9}}, \bibinfo{pages}{387--398}
  (\bibinfo{year}{1971}).

\bibitem{mainardi2012historical}
\bibinfo{author}{F.~Mainardi}, \enquote{\bibinfo{title}{An historical
  perspective on fractional calculus in linear viscoelasticity}},
  \bibinfo{journal}{Fract. Calc. Appl. Anal.} \textbf{\bibinfo{volume}{15}},
  \bibinfo{pages}{712--717} (\bibinfo{year}{2012}).

\bibitem{Rossikhin2010B}
\bibinfo{author}{Y.~A. Rossikhin}, \enquote{\bibinfo{title}{Reflections on two
  parallel ways in the progress of fractional calculus in mechanics of
  solids}}, \bibinfo{journal}{Applied Mech.~Rev.}
  \textbf{\bibinfo{volume}{63}}, \bibinfo{pages}{010701{--}1--010701{--}12}
  (\bibinfo{year}{2010}).

\bibitem{Mainardi2010}
\bibinfo{author}{F.~Mainardi}, \emph{\bibinfo{title}{{Fractional Calculus and
  Waves in Linear Viscoelesticity: An Introduction to Mathematical Models}}},
  \bibinfo{pages}{1--347} (\bibinfo{publisher}{Imperial College Press},
  \bibinfo{address}{London, UK}) (\bibinfo{year}{2010}).

\bibitem{Podlubny1999chapter10_2}
\bibinfo{author}{I.~Podlubny}, \emph{\bibinfo{title}{Fractional differential
  equations}}, chapter \bibinfo{chapter}{10.2} (\bibinfo{publisher}{Academic
  {P}ress}, \bibinfo{address}{New York}) (\bibinfo{year}{1999}).

\bibitem{Treeby2010sectionIIB}
\bibinfo{author}{B.~E. Treeby} and \bibinfo{author}{B.~T. Cox},
  \enquote{\bibinfo{title}{{Modeling power law absorption and dispersion for
  acoustic propagation using the fractional Laplacian}}}, \bibinfo{journal}{J.\
  Acoust.\ Soc.\ Am.} \textbf{\bibinfo{volume}{127}},
  \bibinfo{pages}{2741--2748} (\bibinfo{year}{2010}), \bibinfo{note}{section
  IIB}.

\bibitem{mainardi2011creeprelaxationviscosity}
\bibinfo{author}{F.~Mainardi} and \bibinfo{author}{G.~Spada},
  \enquote{\bibinfo{title}{Creep, relaxation and viscosity properties for basic
  fractional models in rheology}}, \bibinfo{journal}{Eur. J. Phys.}
  \textbf{\bibinfo{volume}{193}}, \bibinfo{pages}{133--160}
  (\bibinfo{year}{2011}).

\bibitem{Rossikhin1997}
\bibinfo{author}{Y.~A. Rossikhin} and \bibinfo{author}{M.~V. Shitikova},
  \enquote{\bibinfo{title}{Applications of fractional calculus to dynamic
  problems of linear and nonlinear hereditary mechanics of solids}},
  \bibinfo{journal}{Appl. Mech. Rev.} \textbf{\bibinfo{volume}{50}},
  \bibinfo{pages}{15--67} (\bibinfo{year}{1997}).

\bibitem{Rossikhin2010}
\bibinfo{author}{Y.~A. Rossikhin} and \bibinfo{author}{M.~V. Shitikova},
  \enquote{\bibinfo{title}{Application of fractional calculus for dynamic
  problems of solid mechanics: Novel trends and recent results}},
  \bibinfo{journal}{Appl. Mech. Rev.} \textbf{\bibinfo{volume}{63}},
  \bibinfo{pages}{010801--1--25} (\bibinfo{year}{2010}).

\bibitem{Bagley83A}
\bibinfo{author}{R.~L. Bagley} and \bibinfo{author}{P.~J. Torvik},
  \enquote{\bibinfo{title}{Fractional calculus --- {A} different approach to
  the analysis of viscoelastically damped structures}}, \bibinfo{journal}{AIAA
  J.} \textbf{\bibinfo{volume}{21}}, \bibinfo{pages}{741--748}
  (\bibinfo{year}{1983}).

\bibitem{Dinzart2009}
\bibinfo{author}{F.~Dinzart} and \bibinfo{author}{P.~Lipinski},
  \enquote{\bibinfo{title}{Improved five-parameter fractional derivative model
  for elastomers}}, \bibinfo{journal}{Arch. Mech.}
  \textbf{\bibinfo{volume}{61}}, \bibinfo{pages}{459--474}
  (\bibinfo{year}{2009}).

\bibitem{Bagley1986}
\bibinfo{author}{R.~L. Bagley} and \bibinfo{author}{P.~J. Torvik},
  \enquote{\bibinfo{title}{On the fractional calculus model of viscoelastic
  behavior}}, \bibinfo{journal}{J. Rheol.} \textbf{\bibinfo{volume}{30}},
  \bibinfo{pages}{133--155} (\bibinfo{year}{1986}).

\bibitem{Glockle1991}
\bibinfo{author}{W.~G. Gl\"ockle} and \bibinfo{author}{T.~F. Nonnenmacher},
  \enquote{\bibinfo{title}{Fractional integral operators and {F}ox functions in
  the theory of viscoelasticity}}, \bibinfo{journal}{Macromolecules}
  \textbf{\bibinfo{volume}{24}}, \bibinfo{pages}{6426--6434}
  (\bibinfo{year}{1991}).

\bibitem{atanakovic2011thermodynamical}
\bibinfo{author}{T.~M. Atanackovic}, \bibinfo{author}{S.~Konjik},
  \bibinfo{author}{L.~Oparnica}, and \bibinfo{author}{D.~Zorica},
  \enquote{\bibinfo{title}{Thermodynamical restrictions and wave propagation
  for a class of fractional order viscoelastic rods}}, \bibinfo{journal}{Abstr.
  Appl. Anal.}  (\bibinfo{year}{2011}), \bibinfo{note}{article {ID} 975694}.

\bibitem{rossikhin2001analysis}
\bibinfo{author}{Y.~A. Rossikhin} and \bibinfo{author}{M.~V. Shitikova},
  \enquote{\bibinfo{title}{Analysis of rheological equations involving more
  than one fractional parameters by the use of the simplest mechanical systems
  based on these equations}}, \bibinfo{journal}{Mech. Time-Depend. Mat.}
  \textbf{\bibinfo{volume}{5}}, \bibinfo{pages}{131--175}
  (\bibinfo{year}{2001}).

\bibitem{Cole1941}
\bibinfo{author}{K.~S. Cole} and \bibinfo{author}{R.~H. Cole},
  \enquote{\bibinfo{title}{Dispersion and absorption in dielectrics {I}.
  {A}lternating current characteristics}}, \bibinfo{journal}{J. Chem. Phys.}
  \textbf{\bibinfo{volume}{9}}, \bibinfo{pages}{341--351}
  (\bibinfo{year}{1941}).

\bibitem{Davis2006}
\bibinfo{author}{G.~B. Davis}, \bibinfo{author}{M.~Kohandel},
  \bibinfo{author}{S.~Sivaloganathan}, and \bibinfo{author}{G.~Tenti},
  \enquote{\bibinfo{title}{The constitutive properties of the brain
  paraenchyma. {P}art 2. {F}ractional derivative approach}},
  \bibinfo{journal}{Med. Eng. Phys.} \textbf{\bibinfo{volume}{28}},
  \bibinfo{pages}{455--459} (\bibinfo{year}{2006}).

\bibitem{Klatt2007}
\bibinfo{author}{D.~Klatt}, \bibinfo{author}{U.~Hamhaber},
  \bibinfo{author}{P.~Asbach}, \bibinfo{author}{J.~Braun}, and
  \bibinfo{author}{I.~Sack}, \enquote{\bibinfo{title}{Noninvasive assessment of
  the rheological behavior of human organs using multifrequency {MR}
  elastography: A study of brain and liver viscoelasticity}},
  \bibinfo{journal}{Phys. Med. Biol.} \textbf{\bibinfo{volume}{52}},
  \bibinfo{pages}{7281--7294} (\bibinfo{year}{2007}).

\bibitem{Kohandel2005}
\bibinfo{author}{M.~Kohandel}, \bibinfo{author}{S.~Sivaloganathan},
  \bibinfo{author}{G.~Tenti}, and \bibinfo{author}{K.~Darvish},
  \enquote{\bibinfo{title}{Frequency dependence of complex moduli of brain
  tissue using a fractional {Z}ener model}}, \bibinfo{journal}{Phys. Med.
  Biol.} \textbf{\bibinfo{volume}{50}}, \bibinfo{pages}{2799--2805}
  (\bibinfo{year}{2005}).

\bibitem{Sack2009}
\bibinfo{author}{I.~Sack}, \bibinfo{author}{B.~Beierbach},
  \bibinfo{author}{J.~Wuerfel}, \bibinfo{author}{D.~Klatt},
  \bibinfo{author}{U.~Hamhaber}, \bibinfo{author}{S.~Papazoglou},
  \bibinfo{author}{P.~Martus}, and \bibinfo{author}{J.~Braun},
  \enquote{\bibinfo{title}{The impact of aging and gender on brain
  viscoelasticity}}, \bibinfo{journal}{NeuroImage}
  \textbf{\bibinfo{volume}{46}}, \bibinfo{pages}{652--657}
  (\bibinfo{year}{2009}).

\bibitem{petrovic2005}
\bibinfo{author}{L.~M. Petrovic}, \bibinfo{author}{D.~T. Spasic}, and
  \bibinfo{author}{T.~M. Atanackovic}, \enquote{\bibinfo{title}{On a
  mathematical model of a human root dentin}}, \bibinfo{journal}{Dent. Mater.}
  \textbf{\bibinfo{volume}{21}}, \bibinfo{pages}{125--128}
  (\bibinfo{year}{2005}).

\bibitem{liu2006}
\bibinfo{author}{J.~G. Liu} and \bibinfo{author}{M.~Y. Xu},
  \enquote{\bibinfo{title}{Higher-order fractional constitutive equations of
  viscoelastic materials involving three different parameters and their
  relaxation and creep functions}}, \bibinfo{journal}{Mech. Time-Depend. Mat.}
  \textbf{\bibinfo{volume}{10}}, \bibinfo{pages}{263--279}
  (\bibinfo{year}{2006}).

\bibitem{Craiem2008}
\bibinfo{author}{D.~O. Craiem}, \bibinfo{author}{F.~J. Rojo},
  \bibinfo{author}{J.~M. Atienza}, \bibinfo{author}{G.~V. Guinea}, and
  \bibinfo{author}{R.~L. Armentano}, \enquote{\bibinfo{title}{Fractional
  calculus applied to model arterial viscoelasticity}},
  \bibinfo{journal}{Latin. Am. Appl. Res.} \textbf{\bibinfo{volume}{38}},
  \bibinfo{pages}{141--145} (\bibinfo{year}{2008}).

\bibitem{Coussot2009}
\bibinfo{author}{C.~Coussot}, \bibinfo{author}{S.~Kalyanam},
  \bibinfo{author}{R.~Yapp}, and \bibinfo{author}{M.~Insana},
  \enquote{\bibinfo{title}{Fractional derivative models for ultrasonic
  characterization of polymer and breast tissue viscoelasticity}},
  \bibinfo{journal}{IEEE Trans.\ Ultrason.\ Ferroelectr.,\ Freq.\ Control}
  \textbf{\bibinfo{volume}{56}}, \bibinfo{pages}{715--725}
  (\bibinfo{year}{2009}).

\bibitem{grahovac2010}
\bibinfo{author}{N.~M. Grahovac} and \bibinfo{author}{M.~Zigic},
  \enquote{\bibinfo{title}{Modelling of the hamstring muscle group by use of
  fractional derivatives}}, \bibinfo{journal}{Comput. Math. Appl.}
  \textbf{\bibinfo{volume}{59}}, \bibinfo{pages}{1695--1700}
  (\bibinfo{year}{2010}).

\bibitem{Metzler2003}
\bibinfo{author}{R.~Metzler} and \bibinfo{author}{T.~F. Nonnenmacher},
  \enquote{\bibinfo{title}{Fractional relaxation processes and fractional
  rheological models for the description of a class of viscoelastic
  materials}}, \bibinfo{journal}{Int. J. Plasticity}
  \textbf{\bibinfo{volume}{19}}, \bibinfo{pages}{941--959}
  (\bibinfo{year}{2003}).

\bibitem{Pritz1996}
\bibinfo{author}{T.~Pritz}, \enquote{\bibinfo{title}{Analysis of four-parameter
  fractional derivative model of real solid materials}}, \bibinfo{journal}{J.\
  Sound.\ Vib.} \textbf{\bibinfo{volume}{195}}, \bibinfo{pages}{103--115}
  (\bibinfo{year}{1996}).

\bibitem{Pritz2001}
\bibinfo{author}{T.~Pritz}, \enquote{\bibinfo{title}{Loss factor peak of
  viscoelastic materials: Magnitude to width relations}}, \bibinfo{journal}{J.\
  Sound.\ Vib.} \textbf{\bibinfo{volume}{246}}, \bibinfo{pages}{265--280}
  (\bibinfo{year}{2001}).

\bibitem{Pritz2003}
\bibinfo{author}{T.~Pritz}, \enquote{\bibinfo{title}{Five-parameter fractional
  derivative model for polymeric damping materials}}, \bibinfo{journal}{J.\
  Sound.\ Vib.} \textbf{\bibinfo{volume}{265}}, \bibinfo{pages}{935--952}
  (\bibinfo{year}{2003}).

\bibitem{sasso2011application}
\bibinfo{author}{M.~Sasso}, \bibinfo{author}{G.~Palmieri}, and
  \bibinfo{author}{D.~Amodio}, \enquote{\bibinfo{title}{Application of
  fractional derivative models in linear viscoelastic problems}},
  \bibinfo{journal}{Mech. Time-Depend. Mat.} \textbf{\bibinfo{volume}{15}},
  \bibinfo{pages}{367--387} (\bibinfo{year}{2011}).

\bibitem{mainardi1994fractional}
\bibinfo{author}{F.~Mainardi}, \enquote{\bibinfo{title}{Fractional relaxation
  in anelastic solids}}, \bibinfo{journal}{Journal of alloys and compounds}
  \textbf{\bibinfo{volume}{211}}, \bibinfo{pages}{534--538}
  (\bibinfo{year}{1994}).

\bibitem{Adolfsson2005}
\bibinfo{author}{K.~Adolfsson}, \bibinfo{author}{M.~Enelund}, and
  \bibinfo{author}{P.~Olsson}, \enquote{\bibinfo{title}{On the fractional order
  model of viscoelasticity}}, \bibinfo{journal}{Mech. Time-Dep. Mater.}
  \textbf{\bibinfo{volume}{9}}, \bibinfo{pages}{15--34} (\bibinfo{year}{2005}).

\bibitem{Chatterjee2005}
\bibinfo{author}{A.~Chatterjee}, \enquote{\bibinfo{title}{{Statistical origins
  of fractional derivatives in viscoelasticity}}}, \bibinfo{journal}{J.\
  Sound.\ Vib.} \textbf{\bibinfo{volume}{284}}, \bibinfo{pages}{1239--1245}
  (\bibinfo{year}{2005}).

\bibitem{Machado2008}
\bibinfo{author}{J.~A.~T. Machado} and \bibinfo{author}{A.~Galhano},
  \enquote{\bibinfo{title}{Fractional dynamics: A statistical perspective}},
  \bibinfo{journal}{J. Comput. Nonlin. Dynam.} \textbf{\bibinfo{volume}{3}},
  \bibinfo{pages}{021201--1--021201--4} (\bibinfo{year}{2008}).

\bibitem{Papoulia2010}
\bibinfo{author}{K.~Papoulia}, \bibinfo{author}{V.~Panoskaltsis},
  \bibinfo{author}{N.~Kurup}, and \bibinfo{author}{I.~Korovajchuk},
  \enquote{\bibinfo{title}{Rheological representation of fractional order
  viscoelastic material models}}, \bibinfo{journal}{Rheol. Acta}
  \textbf{\bibinfo{volume}{49}}, \bibinfo{pages}{381--400}
  (\bibinfo{year}{2010}).

\bibitem{deOliveira2011}
\bibinfo{author}{E.~C. de~Oliveira}, \bibinfo{author}{F.~Mainardi}, and
  \bibinfo{author}{J.~Vaz}, \enquote{\bibinfo{title}{Models based on
  {M}ittag-{L}effler functions for anomalous relaxation in dielectrics}},
  \bibinfo{journal}{Eur. J. Phys.} \textbf{\bibinfo{volume}{193}},
  \bibinfo{pages}{161--171} (\bibinfo{year}{2011}).

\bibitem{bagley1991thermorheologically}
\bibinfo{author}{R.~L. Bagley}, \enquote{\bibinfo{title}{The
  thermorheologically complex material}}, \bibinfo{journal}{Int. J. Eng. Sci.}
  \textbf{\bibinfo{volume}{29}}, \bibinfo{pages}{797--806}
  (\bibinfo{year}{1991}).

\bibitem{bamber2005attenuationandabsorption}
\bibinfo{author}{J.~C. Bamber}, \emph{\bibinfo{title}{Attenuation and
  Absorption}}, chapter~\bibinfo{chapter}{4}, \bibinfo{pages}{93--166}
  (\bibinfo{publisher}{John Wiley \& Sons}, \bibinfo{address}{Chichester, UK})
  (\bibinfo{year}{2005}).

\bibitem{chatelin2012measured}
\bibinfo{author}{S.~Chatelin}, \bibinfo{author}{S.~A. Lambert},
  \bibinfo{author}{L.~Jug\'e}, \bibinfo{author}{X.~Cai}, \bibinfo{author}{S.~P.
  N\"asholm}, \bibinfo{author}{V.~Vilgrain}, \bibinfo{author}{B.~E. Van~Beers},
  \bibinfo{author}{L.~E. Maitre, X.~Bilston}, \bibinfo{author}{B.~Guzina},
  \bibinfo{author}{S.~Holm}, and \bibinfo{author}{R.~Sinkus},
  \enquote{\bibinfo{title}{Measured elasticity and its frequency dependence are
  sensitive to tissue microarchitecture in mr elastography}}, in
  \emph{\bibinfo{booktitle}{Proceedings of the 20th Annual Meeting of ISMRM}}
  (\bibinfo{year}{2012}).

\bibitem{juge2012subvoxel}
\bibinfo{author}{L.~Jug\'e}, \bibinfo{author}{S.~A. Lambert},
  \bibinfo{author}{S.~Chatelin}, \bibinfo{author}{L.~ter Beek},
  \bibinfo{author}{V.~Vilgrain}, \bibinfo{author}{B.~E. Van~Beers},
  \bibinfo{author}{L.~E. Bilston}, \bibinfo{author}{B.~Guzina},
  \bibinfo{author}{S.~Holm}, and \bibinfo{author}{R.~Sinkus},
  \enquote{\bibinfo{title}{Sub-voxel micro-architecture assessment by diffusion
  of mechanical shear waves}}, in \emph{\bibinfo{booktitle}{Proceedings of the
  20th Annual Meeting of ISMRM}} (\bibinfo{year}{2012}).

\bibitem{ODoherty71}
\bibinfo{author}{R.~F. O'Doherty} and \bibinfo{author}{N.~A. Anstey},
  \enquote{\bibinfo{title}{Reflections on amplitudes}},
  \bibinfo{journal}{Geophys. Prosp.} \textbf{\bibinfo{volume}{19}},
  \bibinfo{pages}{430--458} (\bibinfo{year}{1971}).

\bibitem{Garnier2010}
\bibinfo{author}{J.~Garnier} and \bibinfo{author}{K.~S{\o}lna},
  \enquote{\bibinfo{title}{Effective fractional acoustic wave equations in
  one-dimensional random multiscale media}}, \bibinfo{journal}{J.\ Acoust.\
  Soc.\ Am.} \textbf{\bibinfo{volume}{127}}, \bibinfo{pages}{62--72}
  (\bibinfo{year}{2010}).

\bibitem{Meidav1964}
\bibinfo{author}{T.~Meidav}, \enquote{\bibinfo{title}{Viscoelastic properties
  of the standard linear solid}}, \bibinfo{journal}{Geophys. Prospect.}
  \textbf{\bibinfo{volume}{12}}, \bibinfo{pages}{1365--2478}
  (\bibinfo{year}{1964}).

\bibitem{Royer00}
\bibinfo{author}{D.~Royer} and \bibinfo{author}{E.~Dieulesaint},
  \emph{\bibinfo{title}{Elastic Waves in Solids, vol {I}}}
  (\bibinfo{publisher}{Springer}, \bibinfo{address}{Berlin})
  (\bibinfo{year}{2000}).

\bibitem{Atanackovic2002}
\bibinfo{author}{T.~M. Atanackovic}, \enquote{\bibinfo{title}{A modified
  {Z}ener model of a viscoelastic body}}, \bibinfo{journal}{Continuum Mech.
  Therm.} \textbf{\bibinfo{volume}{14}}, \bibinfo{pages}{137--148}
  (\bibinfo{year}{2002}).

\bibitem{Konjik2010}
\bibinfo{author}{S.~Konjik}, \bibinfo{author}{L.~Oparnica}, and
  \bibinfo{author}{D.~Zorica}, \enquote{\bibinfo{title}{Waves in fractional
  {Z}ener type viscoelastic media}}, \bibinfo{journal}{J. Math. Anal. Appl.}
  \textbf{\bibinfo{volume}{365}}, \bibinfo{pages}{259--268}
  (\bibinfo{year}{2010}).

\bibitem{luchko2012fractional}
\bibinfo{author}{Y.~Luchko}, \enquote{\bibinfo{title}{{Fractional wave equation
  and damped waves}}}, \bibinfo{journal}{ArXiv e-prints}
  (\bibinfo{year}{2012}).

\bibitem{widder1971introductionchapter5_13}
\bibinfo{author}{D.~Widder}, \emph{\bibinfo{title}{An Introduction to Transform
  Theory}}, chapter \bibinfo{chapter}{5.13}, Pure and Applied Mathematics
  (\bibinfo{publisher}{Academic Press}) (\bibinfo{year}{1971}).

\bibitem{Vilensky2012}
\bibinfo{author}{G.~Vilensky}, \bibinfo{author}{G.~ter Haar}, and
  \bibinfo{author}{N.~Saffari}, \enquote{\bibinfo{title}{A model of acoustic
  absorption in fluids based on a continuous distribution of relaxation
  times}}, \bibinfo{journal}{Wave Motion} \textbf{\bibinfo{volume}{49}},
  \bibinfo{pages}{93--108} (\bibinfo{year}{2012}).

\bibitem{Pauly1971}
\bibinfo{author}{H.~Pauly} and \bibinfo{author}{H.~P. Schwan},
  \enquote{\bibinfo{title}{Mechanism of absorption of ultrasound in liver
  tissue}}, \bibinfo{journal}{J.\ Acoust.\ Soc.\ Am.}
  \textbf{\bibinfo{volume}{50}}, \bibinfo{pages}{692--699}
  (\bibinfo{year}{1971}).

\bibitem{friedrich1992generalized}
\bibinfo{author}{C.~Friedrich} and \bibinfo{author}{H.~Braun},
  \enquote{\bibinfo{title}{Generalized cole--cole behavior and its rheological
  relevance}}, \bibinfo{journal}{Rheol. Acta} \textbf{\bibinfo{volume}{31}},
  \bibinfo{pages}{309--322} (\bibinfo{year}{1992}).

\bibitem{djordjevic2003cell}
\bibinfo{author}{V.~D. Djordjevi\'c}, \bibinfo{author}{J.~Jari\'c},
  \bibinfo{author}{B.~Fabry}, \bibinfo{author}{J.~J. Fredberg}, and
  \bibinfo{author}{D.~Stamenovi\'c}, \enquote{\bibinfo{title}{Fractional
  derivatives embody essential features of cell rheological behavior}},
  \bibinfo{journal}{Ann. Biomed. Eng.} \textbf{\bibinfo{volume}{31}},
  \bibinfo{pages}{692--699} (\bibinfo{year}{2003}).

\bibitem{Ainslie1998}
\bibinfo{author}{M.~Ainslie} and \bibinfo{author}{J.~G. McColm},
  \enquote{\bibinfo{title}{{A simplified formula for viscous and chemical
  absorption in sea water}}}, \bibinfo{journal}{J.\ Acoust.\ Soc.\ Am.}
  \textbf{\bibinfo{volume}{103}}, \bibinfo{pages}{1671--1672}
  (\bibinfo{year}{1998}).

\bibitem{Bass1995}
\bibinfo{author}{H.~Bass}, \bibinfo{author}{L.~Sutherland},
  \bibinfo{author}{A.~Zuckerwar}, \bibinfo{author}{D.~Blackstock}, and
  \bibinfo{author}{D.~Hester}, \enquote{\bibinfo{title}{{Atmospheric absorption
  of sound: Further developments}}}, \bibinfo{journal}{J.\ Acoust.\ Soc.\ Am.}
  \textbf{\bibinfo{volume}{97}}, \bibinfo{pages}{680--683}
  (\bibinfo{year}{1995}).

\bibitem{Waters2005}
\bibinfo{author}{K.~R. Waters}, \bibinfo{author}{J.~Mobley}, and
  \bibinfo{author}{J.~G. Miller}, \enquote{\bibinfo{title}{Causality-imposed
  ({K}ramers--{K}ronig) relationships between attenuation and dispersion}},
  \bibinfo{journal}{IEEE Trans.\ Ultrason.\ Ferroelectr.,\ Freq.\ Control}
  \textbf{\bibinfo{volume}{52}}, \bibinfo{pages}{822--833}
  (\bibinfo{year}{2005}).

\bibitem{Caputo1967}
\bibinfo{author}{M.~Caputo}, \enquote{\bibinfo{title}{Linear models of
  dissipation whose {Q} is almost frequency independent--{II}}},
  \bibinfo{journal}{Geophys. J. Roy. Astr. S.} \textbf{\bibinfo{volume}{13}},
  \bibinfo{pages}{529--539} (\bibinfo{year}{1967}).

\bibitem{Weaver1981}
\bibinfo{author}{R.~L. Weaver} and \bibinfo{author}{Y.~H. Pao},
  \enquote{\bibinfo{title}{{Dispersion relations for linear wave propagation in
  homogeneous and inhomogeneous media}}}, \bibinfo{journal}{Journ. {M}ath.
  {P}hys} \textbf{\bibinfo{volume}{22}}, \bibinfo{pages}{1909--1918}
  (\bibinfo{year}{1981}).

\bibitem{seredynska2010relaxationdispersion}
\bibinfo{author}{M.~Seredy\'{n}ska} and \bibinfo{author}{A.~Hanyga},
  \enquote{\bibinfo{title}{Relaxation, dispersion, attenuation, and finite
  propagation speed in viscoelastic media}}, \bibinfo{journal}{J. Math. Phys.}
  \textbf{\bibinfo{volume}{51}}, \bibinfo{pages}{092901}
  (\bibinfo{year}{2010}).

\bibitem{weron2000probabilistic}
\bibinfo{author}{K.~Weron} and \bibinfo{author}{A.~Klauzer},
  \enquote{\bibinfo{title}{Probabilistic basis for the {C}ole--{C}ole
  relaxation law}}, \bibinfo{journal}{Ferroelectrics}
  \textbf{\bibinfo{volume}{236}}, \bibinfo{pages}{59--69}
  (\bibinfo{year}{2000}).

\bibitem{nigmatullin2005theoryofdielectric}
\bibinfo{author}{R.~R. Nigmatullin}, \enquote{\bibinfo{title}{Theory of
  dielectric relaxation in non-crystalline solids: from a set of micromotions
  to the averaged collective motion in the mesoscale region}},
  \bibinfo{journal}{Physica B} \textbf{\bibinfo{volume}{358}},
  \bibinfo{pages}{201--215} (\bibinfo{year}{2005}).

\bibitem{stanislavsky2007stochasticnature}
\bibinfo{author}{A.~A. Stanislavsky}, \enquote{\bibinfo{title}{The stochastic
  nature of complexity evolution in the fractional systems}},
  \bibinfo{journal}{Chaos Soliton Fract.} \textbf{\bibinfo{volume}{34}},
  \bibinfo{pages}{51--61} (\bibinfo{year}{2007}).

\bibitem{feldman2012dielectric}
\bibinfo{author}{Y.~Feldman}, \bibinfo{author}{Y.~A. Gusev}, and
  \bibinfo{author}{M.~A. Vasilyeva}, \enquote{\bibinfo{title}{Dielectric
  relaxation phenomena in complex systems}}, \bibinfo{type}{Tutorial},
  \bibinfo{institution}{{K}azan {F}ederal {U}niversity, {I}nstitute of
  {P}hysics} (\bibinfo{year}{2012}).

\bibitem{bhalekar2012generalizedfractional}
\bibinfo{author}{S.~Bhalekar}, \bibinfo{author}{V.~Daftardar-Gejji},
  \bibinfo{author}{D.~Baleanu}, and \bibinfo{author}{R.~Magin},
  \enquote{\bibinfo{title}{Generalized fractional order bloch equation with
  extended delay}}, \bibinfo{journal}{Int. J. Bifurcat. Chaos}
  \textbf{\bibinfo{volume}{22}}, \bibinfo{pages}{1250071--1--1250071--15}
  (\bibinfo{year}{2012}).

\bibitem{berberiansantos2008}
\bibinfo{author}{M.~N. Berberan-Santos}, \bibinfo{author}{E.~N. Bodunov}, and
  \bibinfo{author}{B.~Valeur}, \enquote{\bibinfo{title}{Luminescence decays
  with underlying distributions of rate constants: General properties and
  selected cases}}, in \emph{\bibinfo{booktitle}{Fluorescence of
  Supermolecules, Polymers, and Nanosystems}}, edited by \bibinfo{editor}{M.~N.
  Berberan-Santos} and \bibinfo{editor}{M.~Hof}, volume~\bibinfo{volume}{4} of
  \emph{\bibinfo{series}{Springer Series on Fluorescence}},
  \bibinfo{pages}{67--103} (\bibinfo{publisher}{Springer Berlin Heidelberg})
  (\bibinfo{year}{2008}).

\bibitem{Mittag-Leffler1903}
\bibinfo{author}{M.~G. Mittag-Leffer}, \enquote{\bibinfo{title}{Sur la nouvelle
  fonction ${E}_\alpha(x)$ ({O}n the new function ${E}_\alpha(x)$)}},
  \bibinfo{journal}{C. R. Acad. Sci. Paris} \textbf{\bibinfo{volume}{137}},
  \bibinfo{pages}{554--558} (\bibinfo{year}{1903}).

\bibitem{Wiman1905}
\bibinfo{author}{A.~Wiman}, \enquote{\bibinfo{title}{{\"U}ber den
  {F}undamentalsatz in der {T}heorie der {F}unktionen ${E}_\alpha(x)$ ({A}bout
  the fundamental theorem in the theory of the function ${E}_\alpha(x)$)}},
  \bibinfo{journal}{Acta Mathematica} \textbf{\bibinfo{volume}{29}},
  \bibinfo{pages}{191--201} (\bibinfo{year}{1905}).

\bibitem{Haubold2011}
\bibinfo{author}{H.~J. Haubold}, \bibinfo{author}{A.~M. Mathai}, and
  \bibinfo{author}{R.~K. Saxena}, \enquote{\bibinfo{title}{Mittag-{L}effler
  functions and their applications}}, \bibinfo{journal}{Journal of Applied
  Mathematics} \textbf{\bibinfo{volume}{2011}}, \bibinfo{pages}{1--51}
  (\bibinfo{year}{2011}).

\bibitem{Podlubny1999chapter1-2}
\bibinfo{author}{I.~Podlubny}, \emph{\bibinfo{title}{Fractional differential
  equations}}, chapter \bibinfo{chapter}{1--2} (\bibinfo{publisher}{Academic
  {P}ress}, \bibinfo{address}{New York}) (\bibinfo{year}{1999}).

\bibitem{Djrbashian1966}
\bibinfo{author}{M.~M. Djrbashian}, \emph{\bibinfo{title}{Integral transforms
  and representations of functions in the complex domain}}, chapter
  \bibinfo{chapter}{3--4} (\bibinfo{publisher}{Nauka},
  \bibinfo{address}{Moscow, USSR}) (\bibinfo{year}{1966}), \bibinfo{note}{in
  Russian}.

\bibitem{Djrbashian1993chapter1}
\bibinfo{author}{M.~M. Djrbashian}, \emph{\bibinfo{title}{Harmonic analysis and
  boundary value problems in the complex domain}}, chapter~\bibinfo{chapter}{1}
  (\bibinfo{publisher}{Birkh{\"a}user}, \bibinfo{address}{Basel, Switzerland})
  (\bibinfo{year}{1993}).

\end{thebibliography}


\end{document}